\documentclass[useAMS,usenatbib]{mnras}
\usepackage{graphicx}
\usepackage{graphicx}
\usepackage{amsmath}
\usepackage[singlelinecheck=false]{caption}
\title[IFU spectroscopy of southern PNe IV]
  {IFU spectroscopy of Southern Planetary Nebulae IV: ~~~~~~~A Physical Model for IC 418}
\author[M. A. Dopita et al.]
  {M.A. Dopita,$^{1}$ A. Ali,$^{2,3}$ R.S. Sutherland,$^{1}$ D.C. Nicholls,$^{1}$ \& M. A. Amer,$^{2,3}$ \\
  $^1$Research School of Astronomy and Astrophysics, Australian National University, Cotter Rd., Weston ACT 2611, Australia \\
  $^2$Astronomy Dept, Faculty of Science, King Abdulaziz University, Jeddah, Saudi Arabia \\
  $^3$Department of Astronomy, Faculty of Science, Cairo University, Egypt 
  }
\date{Released 2016 Xxxxx XX}

\pagerange{\pageref{firstpage}--\pageref{lastpage}} \pubyear{2017}

\def\LaTeX{L\kern-.36em\raise.3ex\hbox{a}\kern-.15em
    T\kern-.1667em\lower.7ex\hbox{E}\kern-.125emX}

\begin{document}

\label{firstpage}

\maketitle

\begin{abstract}
 We describe high spectral resolution, high dynamic range integral field spectroscopy of IC418 covering the spectral range 3300-8950{\AA} and compare with earlier data. We determine line fluxes, derive chemical abundances, provide a spectrum of the central star, and determine the shape of the nebular continuum. Using photoionisation models, we derive the reddening function from the nebular continuum and recombination lines. The nebula has a very high inner ionisation parameter. Consequently, radiation pressure dominates the gas pressure and dust absorbs a large fraction of ionising photons. Radiation pressure induces increasing density with radius. From a photoionisation analysis we derive central star parameters; $\log T_{\mathrm eff} = 4.525$K,  $\log L_*/L_{\odot} = 4.029$, $\log g = 3.5$ and using stellar evolutionary models we estimate an initial mass of $2.5 < M/M_{\odot} < 3.0$. The inner filamentary shell is shocked by the rapidly increasing stellar wind ram pressure, and we model this as an externally photoionised shock. In addition, a shock is driven into the pre-existing Asymptotic Giant Branch stellar wind by the strong D-Type ionisation front developed at the outer boundary of the nebula. From the dynamics of the inner mass-loss bubble, and from stellar evolutionary models we infer that the nebula became ionised in the last $100-200$\,yr, but evolved structurally during the  $\sim 2000$\,yr since the central star evolved off the AGB. The estimated current mass loss rate ($\dot M = 3.8\times 10^{-8} M_{\odot}$yr$^{-1}$) and terminal velocity ($v_{\infty} \sim 450$\,km/s) is sufficient to excite the inner mass-loss bubble. While on the AGB, the central star lost mass at $\dot M = 2.1\times 10^{-5} M_{\odot}$yr$^{-1}$ with outflow velocity $\sim 14$\,km/s.
\end{abstract}


\section{Introduction}

Reviewing the literature as listed in the SIMBAD database, we find that a daunting number of articles ($\sim 950$) have studied the compact, young, high surface brightness, and low excitation class planetary nebula (PN) IC 418. These cover the entire accessible electromagnetic spectrum.  Indeed, at first glance, it is difficult to imagine what can be contributed to the discussion on that object that is new.  

The narrow band HST emission line images of this so-called ``Spirograph nebula" show a well-defined oval shape with a simple radial ionization stratification andquasi-regular delicate filamentation. The HST color composite image (H$\alpha$+[O III]) displays a clearly enhanced filamentary [O III] shell emission in the inner bubble region \citep{Sahai11}. These authors suggested that this inner bubble represents emission from very hot gas in the reverse shock generated by the spherical, radiatively driven, fast wind from the central star (CS). Here we present a somewhat different model. To investigate the occurrence of shocks in planetary nebulae (PNe), \citet{Guerrero13} have built [O III]/H$\alpha$ ratio maps which show a decrease at the outer edge in case of IC 418. The optical, near, and mid-IR images probe the presence of different structures around the bright main nebular shell including radial filaments/rays, a system of three concentric rings and two detached haloes \citep{Ramos-Larios12}. Some of these features can be ascribed to enhanced mass-loss during thermal pulses occurring at the end of the AGB phase of stellar evolution \citep{VW93}. Furthermore, they noticed that the progression of the ionization front through the nebula is not homogenous, with the development of instabilities at the outer regions of the ionized shell (which give rise to the ``spirograph" appearance), and the formation of radial structures probably caused by UV photons leaking from the less opaque regions of the ionised shell. 

The central star (CS) of IC 418 (HD 35914) has been classified as an Of (H-rich) type \citep{Acker92} . It has a relatively a low temperature as determined from  the Zanstra ($T_{Z}$(H I) = 34kK) and energy balance (T$_{\rm EB}$ = 36 kK) methods \citep{Pottasch10}. Due to the marginally detected He II $\lambda$4686 emission of IC 418, which (as shown here) seems to originate mainly in the central star , it is hard to derive a realistic He II Zanstra temperature. Presumably, it was this central star HeII emission which led \citet{Phillips03} to derive the much higher He II Zanstra temperature ($T_{Z}$(He II) = 44.5 kK).  

Evidence for a hot shocked stellar wind and possible electron conduction region in the inner region of IC 418 has been found by \citet{Ruiz13} (see the Chandra and HST color composite image of IC 418 in their Figure 3).  These authors find diffuse X-rays in a number of nebulae showing O VI nebular emission in the UV.  In the specific case of IC 418, the \emph{Chandra} data is consistent with thermal emission from a hot plasma confined within the inner [O III] bright filamentary shell at a temperature of 0.26\,keV ( $3\times10^6$K). 

Time-variability structure in the fast winds of the IC 418 CS was detected by \citet{Prinja12}. They reported that the UV resonance line in the IC 418 CS are variable primarily due to the occurrence of blue-ward migrating discrete absorption components.  Polarimetric spectra of the CS indicate mean longitudinal magnetic fields of $\sim 200$ G  \citep{Steffen14}. 

Among recent significant works we note the superb \'echelle spectrum by \citet{Sharpee03, Sharpee04}, which provided accurate line identifications and fluxes over a wide wavelength base, and down to very faint levels. This work enables a comparison of line fluxes with those from our own IFU observations and with those derived from the model which we present in this paper. Both the \citet{Sharpee03} work and the UV and IR spectrophotometry published by \citet{Pottasch04} were used by \citet{Morisset09} to construct a self-consistent stellar and 3D nebular model for IC418 which reproduces the optical and UV stellar observations as well as the nebular IR, deep optical, UV observations, and HST images. The model provides  an effective temperature, $T_{\mathrm eff} = 36.7\pm0.5$ kK and a luminosity of 7700 L$_{\sun}$ for the CS. Further, the model shows that the abundances of O, Ne, and Ar elements are close to solar values, while the elements Si, S, Cl, Mg, and Fe are under-abundant relative to their solar values, indicating trapping of an appreciable fraction of these elements on to dust grains. This work was expanded later by \citet{Escalante12} to show that many of the faint permitted lines of the heavy elements are predominantly excited by fluorescence rather than by recombination.  

\citet{Delgado-Inglada15} have classified IC 418 as a carbon-rich dust PN due to the presence of the infrared broad features at 11 and/or 30 $\micron$ associated with SiC and MgS. From the comparison of He/H, C/O, and N/O abundance ratios derived from the PNe with predictions of nucleosynthesis models, they suggest that PNe with carbon-rich dust descend from stars with masses in the range 1.5-3.0 solar masses. \citet{Otsuka14} confirmed the presence of the broad PAH 3.3 $\micron$ emission band and measured the total flux of the 17.4 and 18.9 $\micron$ emission F(C$_{60}$ - fullerenes) and its fraction with respect to the integrated dust continuum. \citet{Diaz-Luis15} identified 11 diffuse interstellar bands in IC 418, but they found no evidence for the strongest electronic transition of neutral C$_{60}$.  The first detection of the isotope $^{3}$He in IC 418 was reported by \citet{Guzman-Ramirez16}. They derived abundance in the range $1.7\pm0.8\times10^{-3 }$ to $3.8\pm1.7\times10^{-3 }$ for $^{3}$He/H. Such a discovery clearly has impact in the fields of astrophysics and cosmology. 

The integral field unit (IFU) technique (as applied to PNe) was pioneered by \citet{Monreal-Ibero05} and \citet{Tsamis07}. Recently, detailed physical and morpho-kinematical studies using optical IFU data have been obtained by \citet{Danehkar15}, \citet{Danehkar15x} and \citet{Danehkar16} using the Wide Field Spectrograph (WiFeS) instrument \citep{Dopita07,Dopita10} to study the PNe Hen 3-1333, Hen 2-113, Th 2-A and M2-42. The advantages of using the integral field unit (IFU) spectroscopy compared to the long slit spectroscopic technique in the field of planetary nebulae,  were given by \citet{Ali16}. This paper is the fourth in the series examining PNe using the unique capabilities of the WiFeS instrument, which is uniquely well-suited to integral field spectroscopy of compact PNe. In the first paper in this series, \citet{Ali15b} used WiFeS to study the large, evolved and interacting planetary nebula PNG 342.0-01.7, generating an IFU mosaic to cover the full spatial extent of the object. The second paper, \citep{Basurah16} provided a detailed analysis of four highly excited non-type I PNe which casts doubt on the general applicability of the WELS classification. The third paper \citep{Ali16} presented excitation maps, integral field spectroscopy and an abundance analysis of the four PNe: M3-4; M3-6; Hen2-29; Hen2-37. In addition we demonstrated that the CS of M3-6 is another example for the mis-classified WELS group of nebulae. 

These earlier papers used a resolution of $R=7000$ in the red out to 7200\AA\, and a resolution of $R=3000$ in the blue region of the spectrum ($\sim 3600-5600$\AA). In this paper we provide emission line modelling of integral field data covering whole nebula over the full wavelength range of the WiFeS instrument (3300--8950\AA), and at the highest resolution available ($R=7000$). This modelling includes both the effect of the inner [O III] - bright shock, driven into the ionised plasma by the overpressure of the hot shocked stellar wind bubble, and the shock at the outer boundary of the ionised region propagating into the AGB wind, which must be driven by the strong D-type ionisation front formed as both the effective temperature and number of ionising photons produced by the central star rapidly increase with time \citep{Kahn54, Mendis69, Garcia-Segura96}. 

In Section 2, we describe the observations and data reduction, while the nebular and stellar spectra in addition to reddening corrections are discussed in Section 3. The basic parameters of the three zones in a self-consistent model of IC 418, and the details of the model are given in Sections 4 and 5, respectively. Section 6, is dedicated to the results of the nebular model, while Section 7 provides the discussion and conclusions. 

\begin{table*}
 \centering
 \small
   \caption{The log of  WiFeS observations of IC 418}
    \label{Table1}
   \scalebox{0.95}{
  \begin{tabular}{lccccc}
 \hline
   Gratings & No. of  & PA & Exposure  & Date   & Standard \& Telluric Stars \\
& frames &  ($\deg$) & time (s) &  &   \\
   \hline \hline
{\bf IC 418:} & & & & & \\
U7000 \& R7000  &  3 & 90 & 3 & 09/01/2016 & HD 009051, HD 074000  \& HIP 14898 (telluric) \\
U7000 \& R7000  &  3 & 90 & 10 & 09/01/2016 &  " \\
U7000 \& R7000  &  3 & 90 & 30 & 09/01/2016 &  " \\
U7000 \& R7000  &  3 & 90 &  100 & 09/01/2016 &  " \\
{\bf Sky Reference:} & & & & & \\
U7000 \& R7000  &  3 & 0 &  300 & 09/01/2016 &  " \\
{\bf IC 418: } & & & & & \\
B7000 \& I7000 &  6 & 90 & 3 & 09/01/2016 & HD 009051, HD 074000  \& HIP 14898 (telluric) \\
B7000 \& I7000  &  3 & 90 & 10 & 09/01/2016 &  " \\
B7000 \& I7000  &  3 & 90 & 30 & 09/01/2016 &  " \\
B7000 \& I7000  &  3 & 90 & 100 & 09/01/2016 &  " \\
{\bf Sky Reference: } & & & & & \\
B7000 \& I7000  &  3 & 0 &  300 & 09/01/2016 &  " \\
 \hline
 \end{tabular}}
\end{table*}

\section{Observations \& data reduction}\label{Obs}
The integral field spectra of IC 418 were obtained on January 9, 2016 using the WiFeS instrument \citep{Dopita07,Dopita10} mounted on the 2.3-m ANU telescope at Siding Spring Observatory. This instrument delivers a field of view of 25\arcsec $\times$ 38\arcsec at a spatial resolution of either 1.0\arcsec $\times$ 0.5\arcsec or 1.0\arcsec $\times$ 1.0\arcsec, depending on the binning on the CCD. In these observations, we operated in the binned 1.0\arcsec x 1.0\arcsec mode. The data were obtained in the high resolution mode $R \sim 7000$ (FWHM of $\sim 45$ km/s) using all four high-resolution gratings. Observations are made simultaneously in two gratings, as indicated in Table \ref{Table1}. For the U7000 \& R7000 gratings, the dichroic cuts at 480nm (RT480), while for the B7000 \& I7000 gratings, the dichroic cuts at 615nm (RT615). Thus each waveband is observed in a region of high dichroic efficiency, and an adequate overlap in wavelength coverage is obtained between each of the gratings. For details on this, see \citet{Dopita07}.

The wavelength scale was calibrated using the Ne-Ar arc Lamp throughout the night. Arc exposure times are 180s for the  U7000 grating, 100s at B7000, 9s for the  R7000, and 1s for the I7000 grating. Flux calibration was performed using the STIS spectrophotometric standard stars HD 009051 \& HD 074000 \footnote{Available at : \newline {\url{www.mso.anu.edu.au/~bessell/FTP/Bohlin2013/GO12813.html}}}. In addition, a B-type telluric standard HIP 14898 was observed to better correct for the OH and H$_2$O telluric absorption features in the red. The separation of these features by molecular species allows for a more accurate telluric correction by accounting for night to night variations in the column density of these two species. All data cubes were reduced using the PyWiFeS \footnote {\url{http://www.mso.anu.edu.au/pywifes/doku.php.}} data reduction pipeline (\citet{Childress14}). A summary of the spectroscopic observations is given in Table \ref{Table1}. 

The global spectra of each of the objects were extracted from their respective reduced data cubes using a circular aperture matching the observed extent of the PNe using {\tt QFitsView v3.1 rev.741}\footnote{{\tt QFitsView v3.1} is a FITS file viewer using the QT widget library and was developed at the Max Planck Institute for Extraterrestrial Physics by Thomas Ott.}.  This procedure allows for sub-arcsec. differences in the extraction apertures caused by differential atmospheric dispersion. All spectra are sky subtracted using the sky reference exposures listed in Table \ref{Table1}.  In the I7000 and R7000 wavebands, any residual night sky lines were removed using an annular region outside the observed extent of the PNe.

Except in the shortest exposures, the strong lines are saturated on the CCD. Even in the shortest exposure the [O III] $\lambda 5007$\AA\ line is marginally saturated, and for this line the flux was determined from the [O III] $\lambda 4959$\AA\ line, multiplied by its theoretical ratio with respect to [O III] $\lambda 5007$\AA\; $R(5007/4959) = 2.89039$.

The continuum levels and the faint emission lines are determined from the longest exposures in each grating. For the stronger lines, saturation on the CCD is detected by comparison of the measured peak flux between two exposures. If the measured peak flux in the longer exposure is lower, then the line is saturated. The data in the saturated region was then replaced by the unsaturated data in the corresponding region.

\begin{figure}
  \includegraphics[width=\columnwidth]{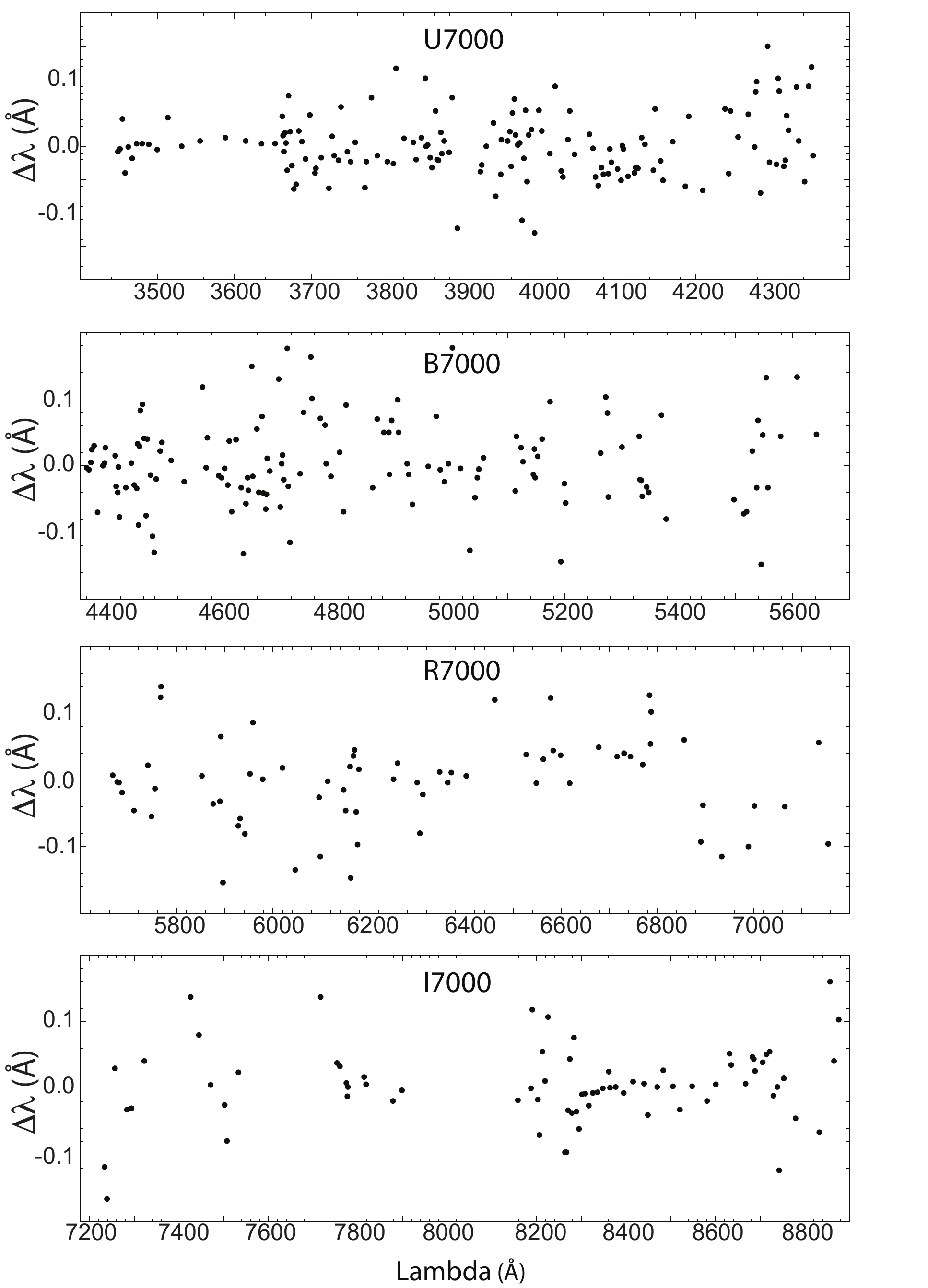}
  \caption{The difference between the measured rest wavelength for emission lines in IC 418 and the NIST wavelength of the line identification as a function of both wavelength and grating used.  The typical RMS error is $\pm 0.05$\AA.} \label{fig1}
 \end{figure}

Finally emission-line fluxes, their uncertainties, the velocity FWHMs and the continuum levels were measured, from the final combined, flux-calibrated spectra, using the interactive routines in {\tt Graf} \footnote{Graf is written by R. S. Sutherland and is available at: {\url {https://miocene.anu.edu.au/graf}} }and in {\tt Lines} \footnote{Lines is written by R. S. Sutherland and is available at: {\url {https://miocene.anu.edu.au/lines}}}. Each line is fit with a single Gaussian, which at this resolution provides a sufficient description of the line profile. The local continuum is fit either side of the line using a linear or a quadratic interpolation. The measured wavelength, the wavelength corrected to rest, and the NIST wavelength of all the identified lines are listed in Table \ref{Table:fluxes} in the Appendix. The mean wavelength error on the identification can be estimated for each grating as a function of wavelength from Fig \ref{fig1}. The identification of the lines in common with this study can be made by comparing  Table \ref{Table:fluxes} with Table 3 of \citet{Sharpee03}. 

\section{Results}
\subsection{Nebular and Stellar Spectra}
\begin{figure*}
  \includegraphics[scale=0.8]{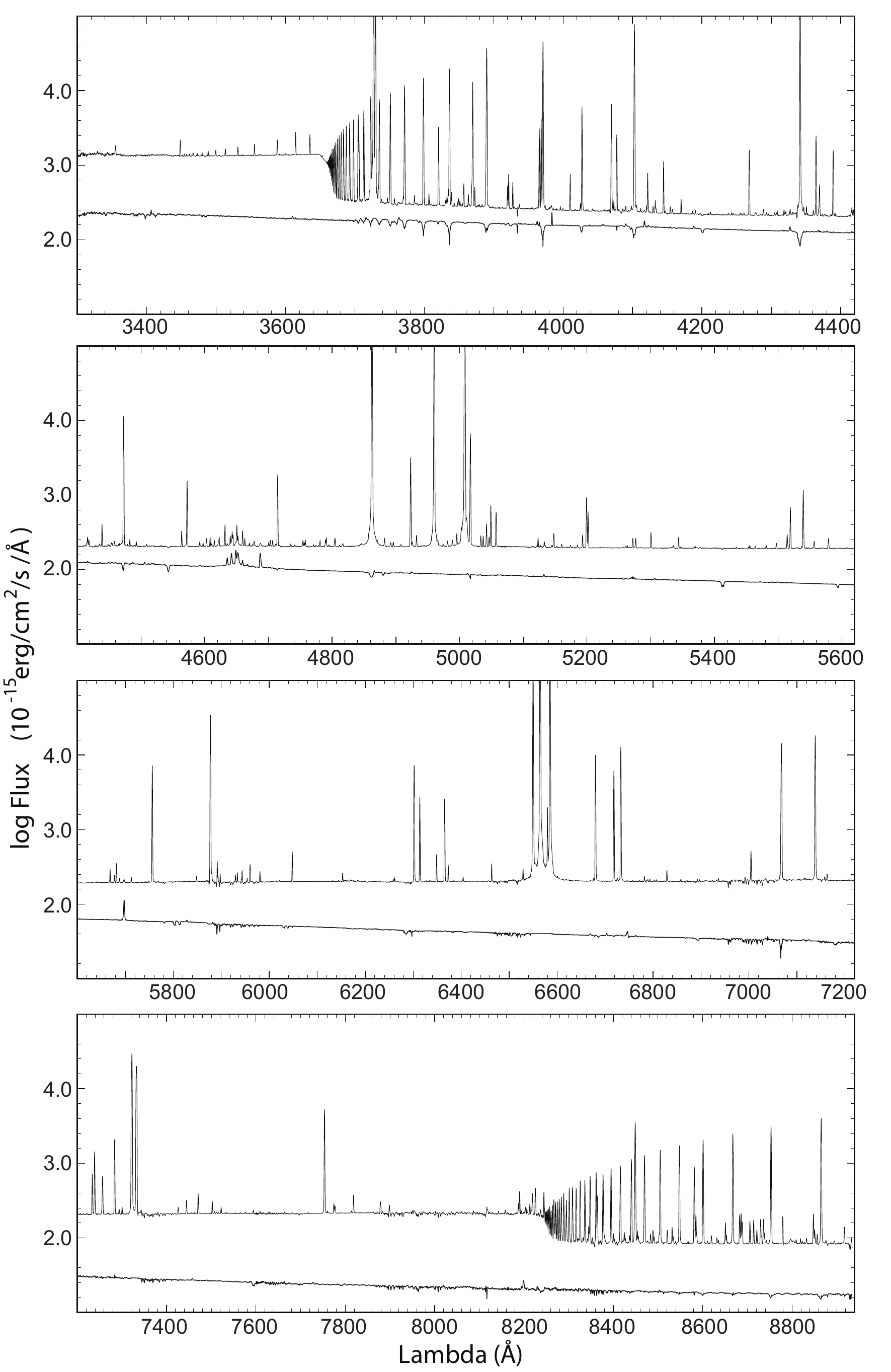}
  \caption{The WiFeS spectrum of the IC 418 nebula (upper curves) and of its central star (lower curves). To bring out fainter lines, the flux scale is logarithmic. For the star, the H$\gamma$ to H$\alpha$ absorption profiles are likely to be in error because of saturation effects.  Note the NIII/CIII/C IV emission line complex in the central star at 4630--4660\AA and the broad He I and H I line absorption lines. The diffuse interstellar absorption bands presented in the high-dispersion spectra by \citet{Diaz-Luis15} at 5780, 5797\AA\ are also visible here.} \label{fig2}
 \end{figure*}

\begin{figure*}
  \includegraphics[scale=0.5]{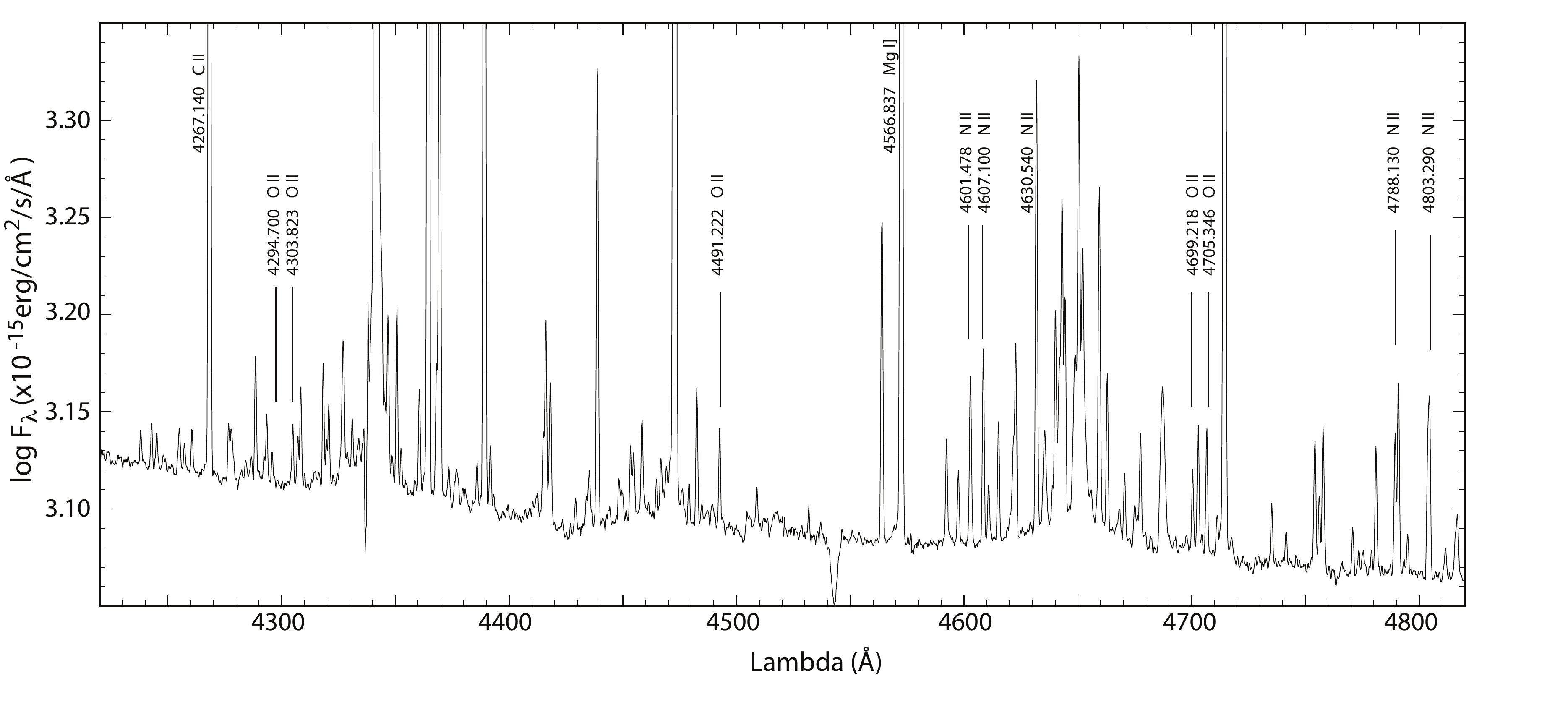}
  \caption{The WiFeS spectrum of the IC 418 nebula enlarged to show a region with O II, NII and CII recombination lines. A few of these are identified on the figure. Note that the noise on the continuum is only $\sim \pm0.02$ dex, and that crowding of faint lines can become a problem for their accurate measurement at this resolution.} \label{fig3}
 \end{figure*}

In Figure \ref{fig2} we show the extracted spectra of the IC 418 nebula, and of its central star, and in Figure \ref{fig3} we show the nebular spectrum in the spectral region 4240-4840\AA\ amplified to bring out the fainter lines and to show the noise in the continuum determination. A number of O II and N II recombination lines are marked. The signal to noise in the nebular continuum is better than 100:1, except in regions of strong residual telluric absorption. These regions show up clearly in the spectrum of the central star. Nonetheless, the quality of the telluric correction is very high, as these residual features only amount to a few percent. Table \ref{Table:fluxes} in the Appendix lists the measured line fluxes with respect to H$\beta = 100$, the estimated flux error and the reddening corrected fluxes (see below for the derivation of these). We also measured line FWHM in velocity, without correction for the instrumental profile ($V_{FWHM} = 45\pm5$\,km/s).

The observed H$\beta$ flux for the full ionised nebula is $\log F_{H\beta} = -9.54\pm0.01$, in agreement with \citet{Pottasch77}.  Previously published values range from -9.52 to -9.71 \citep{Capriotti60, ODell62, Perek71, Kaler73, Carrasco83}. Using a logarithmic reddening correction of $c=0.26$ based on the ratio of H$\alpha$ to H$\beta$, we deduce a reddening corrected flux of  $\log(F_{H\beta}) = -9.28\pm0.01$.

\subsection {Reddening Corrections}
The standard way to infer the reddening correction is from measurement of the Balmer Decrement. On this basis we infer a logarithmic reddening correction of $c=0.26$ for IC 418. This is very similar to the extinction inferred from the $\lambda 2200$\AA\ dip by \citet{Pottasch77} ; $c=0.28$. However, given the high quality spectrophotometric data presented here, we can also use Balmer to Paschen ratios, the recombination lines of He I, and/or the shape of the nebular continuum, \citep{Groves02}. However, to do this, we first need a nebular model against which we can compare our observations. The details of the final model are given below. However, for the purpose of deriving the reddening, we first produced an approximate model to fit the observed line intensities using our estimated logarithmic reddening correction of $c=0.26$. We then inferred an empirical reddening function on the basis of the model line intensities compared with the observations, and using the model nebular continuum compared with the observations. We then used the revised line intensities to obtain an improved model, and repeated the process. Two iterations are sufficient for the process to converge satisfactorily.

\begin{figure}
  \includegraphics[width=\columnwidth]{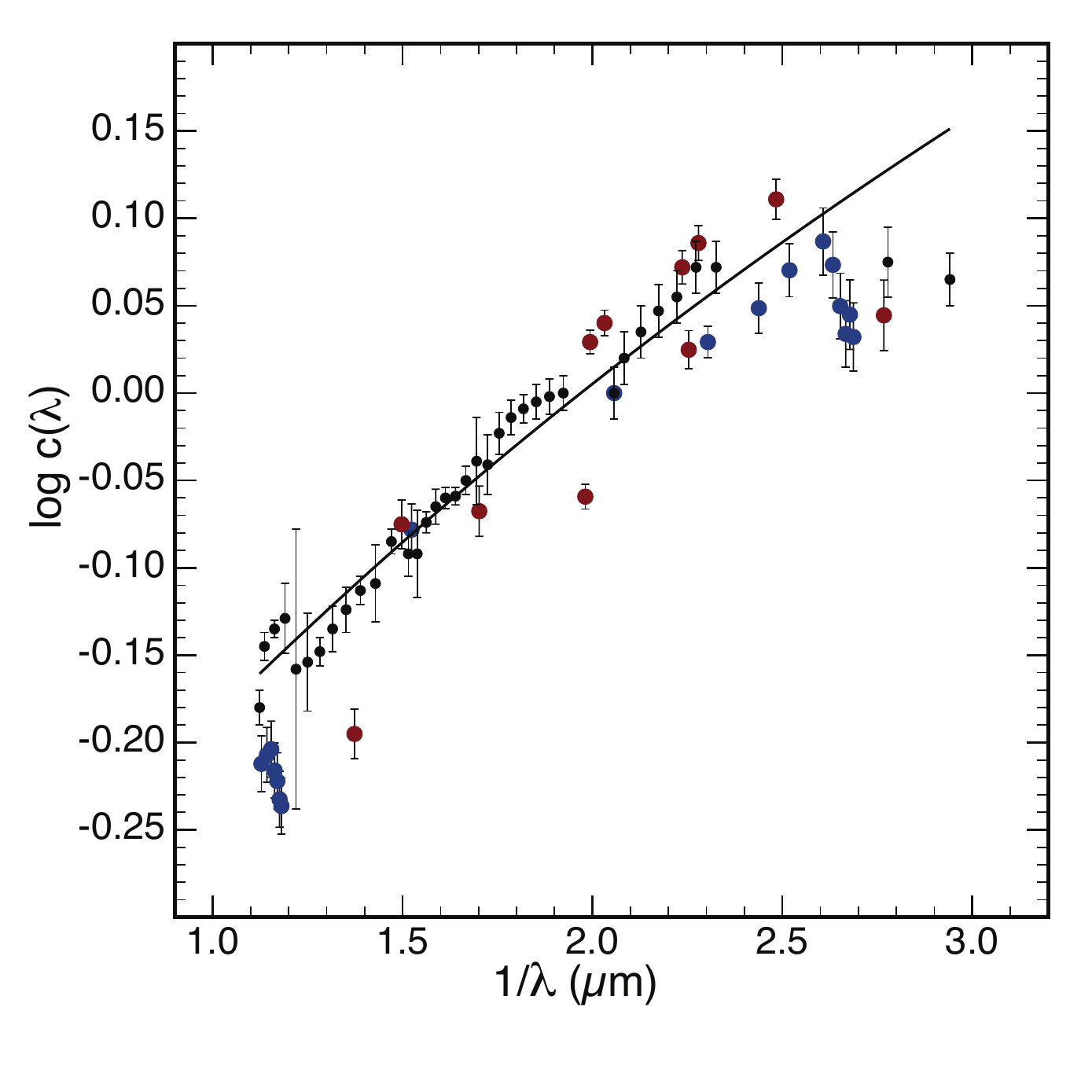}
  \caption{The inferred differential reddening  function in IC 418. The black points are values inferred from the nebular continuum, the blue points are obtained for the Balmer and Paschen lines of Hydrogen while the red points refer to the He I recombination lines. The black line is the adopted fit to the data.} \label{fig4}
 \end{figure}

The resultant differential reddening function is shown in Figure \ref{fig4}. The continuum and the He I lines agree well, as does the Balmer decrement for the well-separated lines. However, the closely spaced Balmer and Paschen lines systematically deviate from the smooth reddening function. A comparison with the \citet{Sharpee03} line intensities suggests that these overlapping lines are not well measured in our line fitting procedure, and should be discounted from the fit.

The derived de-reddened line fluxes are presented in Table \ref{Table:fluxes} in the Appendix. A total of 636 emission lines have been identified.

\subsection{Basic Temperature and Density Diagnostics}
We have used the emission line fluxes measured from the global nebular spectrum, and listed in Table \ref{Table:fluxes} in the Appendix to determine electron temperatures, densities and ionic abundances using the Nebular Empirical Abundance Tool (NEAT; \citet{Wesson12} to derive the electron temperature and density from the low and medium-ionization zones. The NEAT code use the Monte Carlo technique to propagate the statistical uncertainties from the line fluxes to the temperature, density, ionic, and elemental abundances.  The NEAT temperatures and electron densities are given in Table \ref{Table:TemDen}, where we also include the values derived from our detailed photoionisation modelling described below as well as those previously derived in the literature.

In Table \ref{Table:ICFs} we provide the derived elemental abundances from this work, and from the literature. Note that the optical recombination lines (ORLs) of both N and O give a much larger abundance than the coliisionally excited lines (CELs). This difference has been shown by \citet{Escalante12} to be mainly caused by fluorescent effects caused by the UV continuum of the central star, and is discussed in detail in Section \ref{Sec6.3} below.

\begin{table*}
\centering \caption{Average electron temperatures and densities of IC 418 derived from the IFU spectrum compared with those given in the  literature, and by the self-consistent photoionisation model described below.} \label{Table:TemDen} \scalebox{1.0}{
\begin{tabular}{|p{2.0cm}|p{1.2cm}|p{1.2cm}|p{1.2cm}|p{1.2cm}|p{1.2cm}|p{1.2cm}|p{1.2cm}|p{1.2cm}|p{1.2cm}|}
 \hline
      {\bf Object }        & \multicolumn{5}{|c|}{\bf Temperature (K)} & \multicolumn{3}{|c|}{\bf Density (cm$^{-3}$)} \\
               & $T_{[\mathrm {O III]}}$ & $T_{[\mathrm {N II]}}$ &  $T_{\mathrm {[S II]}}$ & $T_{\mathrm {[Ar III]}}$  & $T_{\mathrm{[O II]}}$ & $n_{\mathrm{[S II]}}$ & $n_{\mathrm {[O II]}}$ & $n_{[\mathrm{[Cl III]}}$ \\
                \hline
{\bf IC 418}  &   &    &    &     &    &  &   &  \\
Observed:      & 8775$^{+75}_{-75}$  & 10008$^{+420}_{-420}$  &  9879$^{+4941}_{-4878}$ & 8489$^{+108}_{-108}$ & 9216$^{+150}_{-190}$  & 8374$^{+5024}_{-6382}$    & 11874$^{+8078}_{-3688}$   & 11178$^{+679}_{-679}$  \\
Model: & 8800    & 8830  &  8680 &  8879  & 8856 & 11950  & 12120   & 11450 \\
 &   &    &    &     &    &  &   &  \\
Ref (1) &  8780$^{+150}_{-190}$   & 9530$^{+390}_{-370}$ &    &  9100   &  & 15300$^{+16000}_{-6600}$       & 9000$^{+4800}_{-2300}$   &  10400$^{+1700}_{-1800}$ &  \\
 &   &    &    &     &    &  &   &  \\
Ref (2) &   8900$^{+400}_{-400}$   & 9400$^{+900}_{-1400}$  & 7000$^{+4000}_{...}$   &  9000$^{+500}_{-400}$    &  10000$^{+4000}_{-3000}$     &   17000$^{...}_{-9000}$    & 10000$^{+17000}_{-5000}$  & 11000$^{+4000}_{-2000}$ \\
 &   &    &    &     &    &  &   &  \\
Ref (3) &   9100    & 9400  &    &  9100   &  & 3100       & 5000  &  4300 &  \\
 \hline
\end{tabular}}
\begin{minipage}[!t]{16cm}
{\small {\bf References:} (1) \citet{Delgado-Inglada15}; (2)
\citet{Sharpee04};   (3) \citet{Pottasch04}}
\end{minipage}
\end{table*}

\begin{table*}
\centering \caption{Observed abundances by number with respect to H both derived from the NEAT code
and compared with previous work.} \label{Table:ICFs} \scalebox{1.0}{
\begin{tabular}{lcccc}
 \hline
  & \multicolumn{2}{|c|}{\bf Abundances from NEAT} & \multicolumn{2}{|c|}{\bf Abundances from literature}\\\cline{2-3} \cline{4-5} \\

Element & ORLs & CELs & Ref 1 & Ref 2 \\
\hline 
\vspace{0.2cm}
He/H & 7.97E-2$^{+1.9E-3}_{-1.9E-3}$ & -- & 9.1E-02$^{+9.1E-4}_{-9.1E-4}$& $>$7.20E-2 \\
\vspace{0.2cm}
C/H & 5.50E-4$^{+1.1E-5}_{-1.1E-5}$ & -- & 6.17E-4$^{+1.9E-4}_{-2.1E-4}$ & 6.20E-4 \\
\vspace{0.2cm}
N/H &  3.10E-4$^{+4.7E-6}_{-4.7E-6}$ & 4.51E-5$^{+3.4E-6}_{-4.5E-5}$ & 6.61E-5$^{+1.1E-5}_{-7.3E-6}$ & 9.50E-5 \\
\vspace{0.2cm}
O/H & 9.97E-4$^{+8.1E-4}_{-3.0E-4}$ & 2 .20E-4$^{+6.6E-5}_{-2.2E-4}$ & 3.47E-4$^{+3.1E-5}_{-3.1E-5}$ & 3.50E-4 \\
\vspace{0.2cm}
Ne/H & 7.32E-5$^{+2.2E-5}_{-1.6E-5}$ &3.16E-5$^{+8.6E-6}_{-4.0E-6}$ & 3.47E-5$^{+1.7E-6}_{-4.2E-6}$ & 8.8E-5\\
\vspace{0.2cm}
Ar/H & -- &1.72E-6$^{+5.3E-7}_{-1.7E-6}$ &  1.12E-6$^{+1.6E-7}_{-3.5E-7}$  & 1.80E-6 \\
\vspace{0.2cm}
S/H & -- &1.89E-6$^{+2.1E-7}_{-1.1E-7}$ & -- & 4.4E-6 \\
\vspace{0.2cm}
Cl/H & -- & 8.92E-8$^{+3.1E-9}_{-2.9E-9}$ &   8.7E-8$^{+3.5E-9}_{-9.6E-9}$ & -- \\
\vspace{0.2cm}
N/O &    0.31 &  0.21      &    0.19     &  0.27   \\
 \hline
\end{tabular}}
\begin{minipage}[!t]{16cm}
{\small {\bf References:} (1) \citet{Delgado-Inglada15}; (2)
\citet{Pottasch04}}
\end{minipage}
\end{table*}

\newpage
\section{A physical model for IC418}
We aim to construct a self-consistent model for IC 418 consisting of three separate zones:
\begin{enumerate}
\item{An inner photoionised shock driven by the accelerating stellar wind of the central star, }
\item{A photoionised nebular shell, and}
\item{An outer shock in the AGB wind driven by the over-pressure of the strong D-Type ionisation front.}
\end{enumerate}
Here, we estimate the basic parameters of these three zones.

\subsection{The inner stellar wind bubble shock}\label{4.1}
The inner photoionised shocked shell is expected to exist from the theory of mass-loss bubbles \citep{Dyson72, Weaver77, Schmidt-Voigt87a, Schmidt-Voigt87b, Marten91} and is clearly visible in the HST \emph{Hubble Heritage} image (\url{http://heritage.stsci.edu/2000/28/big.html}) as an elliptical filamentary ring, enhanced in [O III] emission. This ring, approximately 0.01\,pc in diameter,  is coincident with the extent of the diffuse X-ray emission seen with \emph{Chandra} \citep{Ruiz13}. The X-ray emission arises from a thermal plasma at $T_e \sim 3\times10^6$K. This temperature would be produced by a shocked stellar wind from the central star, provided that the terminal velocity of this wind is $v_w \sim 500$\,km/s -- or somewhat faster if cooling of the hot plasma by thermal conduction is important. This compares with the value of $v_w$ estimated from the UV observations of the central star by \citet{Morisset09}; 450\,km/s.

The pressure driving the outer stellar wind shock can be estimated from the properties of the stellar wind itself. From \citet{Morisset09}, the stellar mass-loss rate $\dot M = 3.8\times 10^{-8} M_{\odot}$yr$^{-1}$. The momentum flux in the stellar wind is converted to thermal pressure at the edge of the free-wind region, which theoretical models indicate lies between 0.5--0.7 of the radius of the shock in the swept up AGB wind. At a distance of 1.0\,kpc, derived below, this [O III] - bright stellar wind shock lies at a radius of $\sim 0.01$\,pc. Using a stellar wind velocity of $v_w \sim 500$\,km/s, we find that both the hot X-ray bubble and the shocked [O III] bubble has $\log P/k = 8.2-8.5$\,cm$^{-3}$K. We will use this estimate later to compute the velocity of the stellar wind bubble shock in the photoionised nebula.

\subsection{The photoionised nebula}\label{4.2}
A self-consistent model of the photoionised region of IC 418 has been presented by  \citet{Morisset09}. In order to reproduce the surface brightness distribution, a particular density distribution was imposed for both the inner photoionised stellar wind shock and for the photoionised nebula as a whole. However, this distribution may be a natural consequence of the somewhat unusual parameters of the nebula. Based on an inner radius of 0.01\,pc, and depending upon the density at this inner radius, we estimate that the ionisation parameter at the inner boundary of the photoionised nebula is in the range $-0.1 > \log{\cal U} > -0.5$. This puts IC 418 very firmly into the category of nebulae in which radiation pressure exceeds the gas pressure \citep{Dopita02, Dopita06, Davies16}. In this regime, dust competes very successfully with the ionised plasma to absorb photons in the Lyman continuum, reducing the extent of the photoionised nebula \citep{Dopita03}, while at the same time, radiation pressure induces a steeply increasing density gradient though the photoionised nebula, as the radiation field is absorbed and its pressure adds to the initial gas pressure. Such a density gradient is observed in IC 418, and our modelling seeks to discover whether this is simply due to the action of radiation pressure.

\subsection{The outer shock in the AGB wind}\label{4.3}
As a consequence of the pressure in the photoionised region, which includes both the gas pressure, and the pressure associated with the absorbed fraction of the radiation field at the outer ionised boundary, a strong D-type ionisation front is formed  \citep{Kahn54, Mendis69, Garcia-Segura96}. The structure of this is as follows. Immediately beyond the photoionised region, lies a strongly compressed un-ionised shell. The pressure in this shell matches the sum of the gas pressure and radiation pressure in the ionised region, plus the pressure associated with the recoil momentum of the newly-ionised gas flowing into the photoionised region. Because this un-ionised shell has a much higher pressure than the surrounding undisturbed AGB wind, an expanding shock (assumed isothermal) is located at its outer boundary, such that the ram pressure of the AGB wind being swept up by the shock is equal to the internal pressure of the shell.

The dense shell of un-ionised gas around the strong D-Type ionisation front is Rayleigh-Taylor unstable \citep{Frieman54, Spitzer54}. However, even if the shell is accelerating, these instabilities can be stabilised by recombination \citep{Kahn58, Axford64, Williams99}, and in this case, the ionisation front may oscillate about its mean instantaneous radius with time \citep{Mizuta05}. A full 3-D treatment for the specific case of an ionisation front propagating into a $r^{-2}$ density distribution (which is particularly applicable to PNe) has been made by \citet{Whalen08}. This work shows that fine perturbations or crinkles in the D-Type ionisation front grow, merge and ultimately break out to form ``elephant trunks'' along with isolated neutral inclusions remaining from the dense shell. It would appear that the ``spirograph'' appearance of the outermost parts of IC 418 are an early manifestation of this ionisation front instability. 

The velocity of the outer shock in the AGB wind can be estimated by combining the angular expansion rate measured by \citet{Guzman09} with the HI absorption velocity determined by \citet{Taylor87}. Adopting a distance of $1.0\pm0.1$\,kpc derived from the model presented below, and the angular expansion rate of the ionization front in IC 418 from \citet{Guzman09} ($5.8 \pm1.5$\,mas/yr), gives a shell expansion velocity of 27.4\,km/s. However, the H I absorption feature detected by  \citet{Taylor87} is at -13.2\,km/s with respect to the systemic velocity. This implies that the outer shock has a velocity equal to the difference between these numbers; $\sim14$\,km/s. This is strongly supersonic with respect to the cool AGB wind, and such a shock is still capable of producing some optical emission.

\section{Details of the Model}

\subsection{The Photoionised Nebula}
To model IC 418, we have used the {\tt Mappings V version 5.1.12} code (Sutherland et al. 2017, submitted) {\footnote{Available at {\url{https://miocene.anu.edu.au/mappings}}}. Earlier versions of this code have been used to construct photoionisation models of H\,II regions, PNe, Herbig-Haro Objects,
supernova remnants and narrow-line regions excited by AGN.  This code is the latest version of the {\tt Mappings 4.0} code earlier
described in \citep{Dopita13}, and includes many upgrades to both the input atomic physics and the methods of solution.

We choose as the initial abundance set  the local galactic concordance (LGC) abundances \citep{Nicholls17} based upon the
\citet{Nieva12} data on local galactic main sequence B stars. This provides the abundances of the main coolants, H, He, C, N. O, Ne, Mg,
Si and Fe and the ratios of N/O and C/O as a function of abundance. For the light elements we use the \citet{Lodders09} abundance, while
for all other elements the abundances are based upon \citet{Scott15a, Scott15b} and \citet{Grevesse10}. The individual elemental abundances are then iterated from this initial set in order to minimise the offset of the model with respect to the observations for all the lines of any given element.

In the modelling, dust physics is very important. Not only do dust grains remove coolants from the nebular gas, but they are also an important source of photoelectric heating \citep{Dopita00}. Furthermore, the pressure of the radiation field is primarily coupled to the dust absorption, and in objects such as IC 418, the effect of competition by dust grains with the gas for the absorption Lyman continuum photons is fierce. IC 418 shows a large depletion of the heavy elements onto dust, as evidenced by the depletion of Ca, Mg, and Fe (see below). This suggests the presence of silicates, since we have no physical mechanism to deplete these elements in the stellar envelope.

The detailed dust physics as currently implemented in the Mappings V code has been described in detail by \citet{Dopita05}. Briefly, this consists of PAHs (not included in the IC 418 model), amorphous carbon and silicate grains. For the latter two categories, we use 80 bins in the size distribution, usually taken as a standard \citet{MRN} distribution.  We allow for grain charging, photoelectron emission and grain temperature fluctuations in the computation of the IR emission. An improved version of this dust physics has been implemented by \citet{Morisset12} for the specific case of IC 418, and these authors succeed in reproducing the IR spectrum of the dust thermal emission. As a consequence, we do not attempt to emulate this work here. For the initial depletion factors we use the \citet{Jenkins09} scheme, with a base depletion of Fe of  2.25\,dex, but these are adjusted by individual element to best fit the observations. We obtain a final dust to gas mass ratio of $1.12\times10^{-2}$. The carbon rich nature of the nebula naturally gives rise to a carbon grain dominant dust composition in the models. The dust to gas mass ratio for the carbon grains alone is $3.8\times10^{-3}$. This should be compared to the value of $\sim 6\times10^{-4}$ derived by  \citet{Hoare90} and \citet{Meixner96}. It should be noted that these works provide a reasonable fit to the cool dust emission, but fail by a factor of 3 or so to explain the hot dust emission seen by \citet{Omont95} which peaks at $\sim 30\mu$m, and which is ascribed by these latter authors as possibly due to MgS.

For the central star, we follow \citet{Morisset09} in using the {\tt{CMFGEN}} model atmospheres from \citet{Hillier98}. \citet{Morisset09} used the detailed profiles of the hydrogen absorption lines to constrain the effective gravity of the star ($\log g = 3.5$) and used the excitation of the nebula to constrain the effective temperature ($T_{\mathrm eff} = 36.7$kK). Here we have iterated the effective temperature to match the observed excitation of the nebula, once the density structure has been fixed to reproduce the density sensitive line ratios, with the exception of [Ar IV] which is produced in the inner photoionised shock produced by the stellar wind. We find $T_{\mathrm eff} = 33.5$kK, somewhat lower than that obtained by \citet{Morisset09}, probably because of the different nebular dust physics used here. As mentioned above, for this nebula the radiation pressure acting mainly on dust produces a strong positive radial density gradient. In our models, at the inner radius of 0.01\,pc., the ionisation parameter is $\log {\cal U} = -0.1$ and the gas pressure is $\log P/k = 7.4$. As the radiation field is absorbed in the ionised shell, the pressure rises to $\log P/k = 8.4$\,cm$^{-3}$K.  As a result, the inner electron density is only $n_e =1550$\,cm$^{-3}$, but it peaks at $n_e =27160$\,cm$^{-3}$. This causes the steep increase in surface brightness empirically modelled by \citet{Morisset09}. This is discussed further below.

Although the nebula is mildly elliptical (outer diameter $10.7\times13.7$\,arc sec. and inner diameter  $3.2 \times4.3$\,arc sec.), we model it as a spherical shell with filling factor unity. Our model aims to match both the observed mean diameter and the absolute H$\beta$ flux at the assumed distance. Because of the competition of dust for the ionising photons, the method used by \citet{Basurah16} to estimate the distance does not work. For a given luminosity of the central star, the computed absolute H$\beta$ luminosity hardly changes with distance, because, for smaller assumed distances, the ionisation parameter at the inner radius of the shell rises, and the fraction of ionising photons absorbed by dust rises. However, a solution which matches both the observed mean diameter and the absolute H$\beta$ flux can still be obtained. With this we find $\log L_* = 4.029$ and a distance, $D=1.0\pm0.1$\,kpc. At assumed distances of 0.9, 1.0 and 1.1\,kpc, the corresponding computed absolute H$\beta$ fluxes are $\log L_{H\beta} = 34.76, 34.78$ and 34.80. The observed absolute H$\beta$ fluxes corrected for reddening and using these same assumed distances are $\log L_{H\beta} = 34.69, 34.78$ and 34.87, respectively. 

In order to measure the goodness of fit of any particular photoionisation model, we measure the degree to which it reproduces the density-sensitive line ratios, and we also seek to minimise the L1-norm for the fit for the main coolant lines, and for the H and He recombination lines - for a total of 34 lines. That is to say that we measure the modulus of the mean logarithmic difference in flux (relative to H$\beta$) between the model and the observations \emph{viz.};
\begin{equation}
{\rm L1} =\frac{1}{m}{\displaystyle\sum_{n=1}^{m}} \left | \log \left[ \frac{F_n({\rm model})} {F_n({\rm obs.)}} \right]  \right |. \label{L1}
\end{equation}
This procedure weights fainter lines equally with stronger lines, and is therefore more sensitive to the values of the input parameters of the model. Once a satisfactory best-fit is obtained (${\rm L1} = 0.051$), we adjust the abundances of the species which are unimportant in the thermal balance of the nebula, or for species for which only one or two emission lines are observed in the optical.

A detailed description of the photoionisation model fit is deferred until later, in order to also include the shocked zones of the nebular model.

\subsection{The inner stellar wind bubble shock}
In section \ref{4.1} we estimated that the driving pressure of the shock driven by the X-ray bubble into the photoionised PNe material lies in the range $\log P_s/k = 8.2-8.5$\,cm$^{-3}$K. However, in the previous section, we found that the pre-shock pressure in the photoionised shell is only  $\log P_{PNe}/k = 7.4$\,cm$^{-3}$K, and the total hydrogen particle density is $n_H = 1400$\,cm$^{-3}$. Adopting  $\log P_s/k = 8.4$\,cm$^{-3}$K, we find that the resultant ram pressure can drive a $40\pm5$\,km/s (Mach number = 3) shock, which heats the immediate post-shock gas to $T_e = 30400$\,K. As the gas cools, it is compressed, and reaches a computed electron density of $n_e = 21800$\,cm$^{-3}$ where it is in equilibrium with the stellar radiation field. The computed equilibrium temperature is $T_e = 8840$\,K, somewhat higher than the equilibrium temperature in the inner photoionised shell ($T_e \sim 8300$\,K). From the model, the time taken for the shocked gas to reach  photoionisation equilibrium is only $\sim 10$\,yr.

The H$\beta$ flux from the shocked shell is only a small fraction of the total luminosity of the PNe. This fraction is roughly proportional to the mean age of the shocked shell. For an assumed age of 100\,yr, it only accounts for $\sim 1.0$\% of the total H$\beta$ flux. However, this shock contributes appreciably to the [O III] emission, since its [O III] $\lambda 5007$/H$\beta$ ratio is 11.44. The shock contributes even more to the highest excitation lines. For example, the [Ne III]$\lambda3869$/H$\beta$ ratio in the shocked shell is 0.66, while in the nebula as a whole it is only 0.024. This shock contribution to the high excitation lines is a further reason why we derive a lower effective temperature for the central star than \citet{Morisset09}.

The inner shock should manifest itself as a region of enhanced electron temperature. This is seen clearly in the WiFeS [O III]$\lambda\lambda4363/4959$ line ratio map; see Figure \ref{fig5}.

\begin{figure}
 \includegraphics[width=\columnwidth]{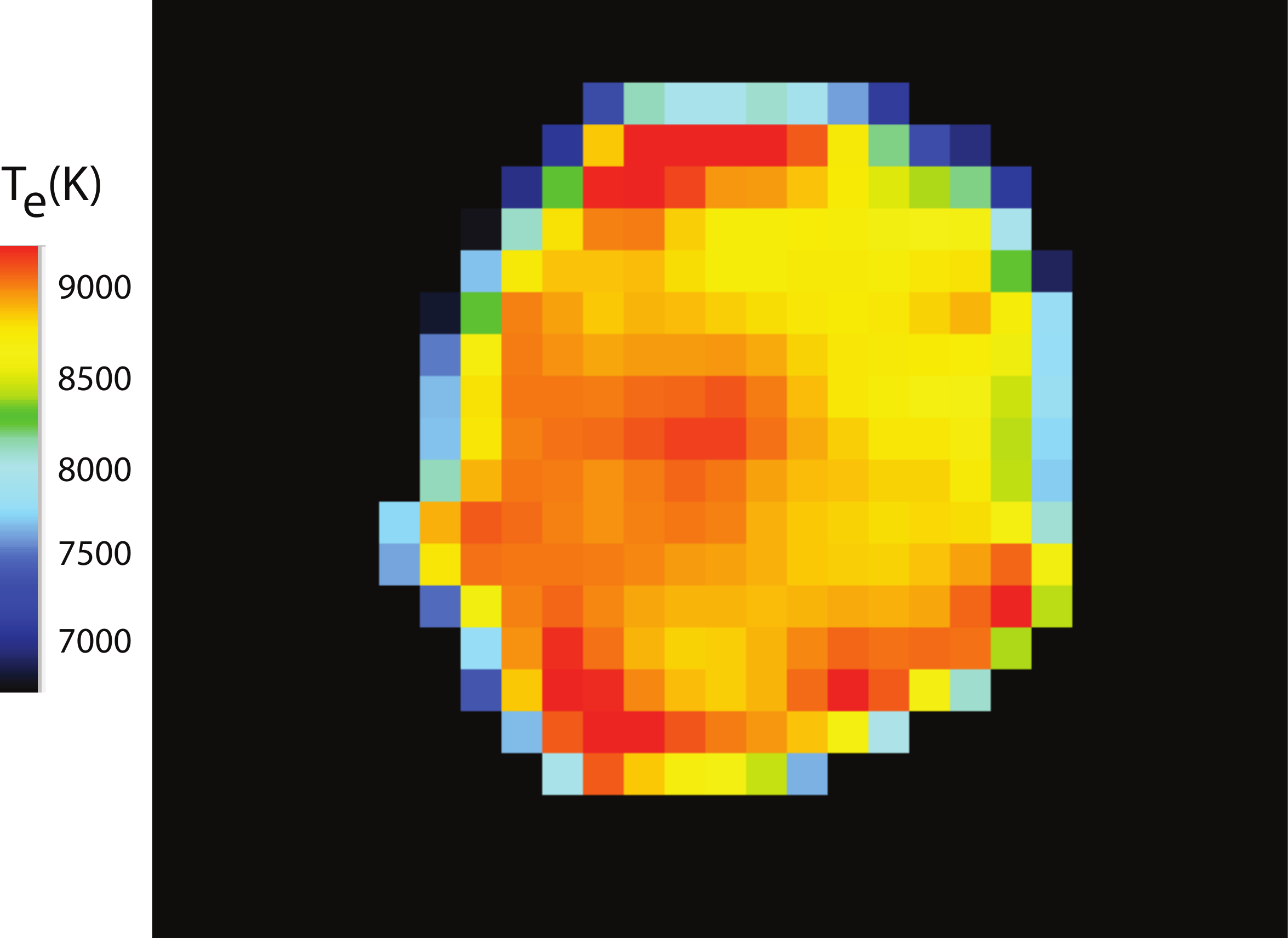}
 \caption{The [O III] $\lambda\lambda4363/4959$ line ratio map converted to mean temperature along the line of sight. The WiFeS image covers $27\times22$\,arc sec. Enhanced temperature associated with the inner shock can be clearly seen, as well as a radial temperature gradient followed by a steep decline at the boundary of the [O III] - emitting zone. This is in qualitative accord with the models which predict a temperature of $T_e = 8840$\,K in the inner shocked shell, an inner temperature of $T_e = 8300$\,K in the photoionised region rising to a peak of $T_e = 9050$\,K before declining to $T_e = 8630$\,K at the boundary of the [O III] - emitting zone.} \label{fig5}
\end{figure}

\subsection{The outer shock in the AGB wind}
The outer shock is a very slow shock proceeding into a medium with a low degree of ionisation. As such, it bears some resemblance to the Herbig-Haro shocks modelled by Dopita \& Sutherland (2017, in press). Such shocks produce much enhanced [O I] and [N I] emission, and also enhanced [S II] emission. We model the shock with a velocity of 14\,km/s, derived from the arguments presented in Section \ref{4.3}. The post-shock pressure is  $\log P/k = 8.4$, which, with this shock velocity implies a pre-shock density of  $n_H = 10000$\,cm$^{-3}$, at a fractional ionisation of a few percent. 

The resulting shock is very feeble in its optical emission, accounting to only about 1.0\% of the H$\beta$ emission of the photoionised shell. However, the computed [N I] $\lambda\lambda 5198,5200$/H$\beta$ ratio is high, about 0.6, as is the [O I] $\lambda 6300$/H$\beta$ ratio; $\sim 1.5$. As a consequence, the shock provides more than 50\% of the total flux of these lines in the model. This accounts for the shortfall in [N I] and [O I] noted by the authors in the pure photoionisation model of \citet{Morisset09}.

\subsection{Overall Structure of Model}
The overall structure of the resulting model is shown in Figure \ref{fig:neTe}, where we show the run of temperature and electron density as a function of radius. The inner shocked and photoionised shell is very thin with the post-shock temperature excursion appearing almost as a delta function. The steep rise in electron density though the main nebula is very evident, although the density variations are more subtle. The outer boundary of the [O III] --emitting zone is marked by small discontinuities in both temperature and density. The boundary of the Str\"omgren sphere is marked by a sharp drop in temperature and electron density. The outer shock produces a spike in both temperature and electron density.

InTable \ref{table:shocks} we show the relative contributions of the inner and outer shocks to the total line fluxes of some key lines. These lines are the most sensitive to the presence of the shocks. The change in relative intensity for the model with and without shocks is at most a factor of two.

\begin{figure}
 \includegraphics[width=\columnwidth]{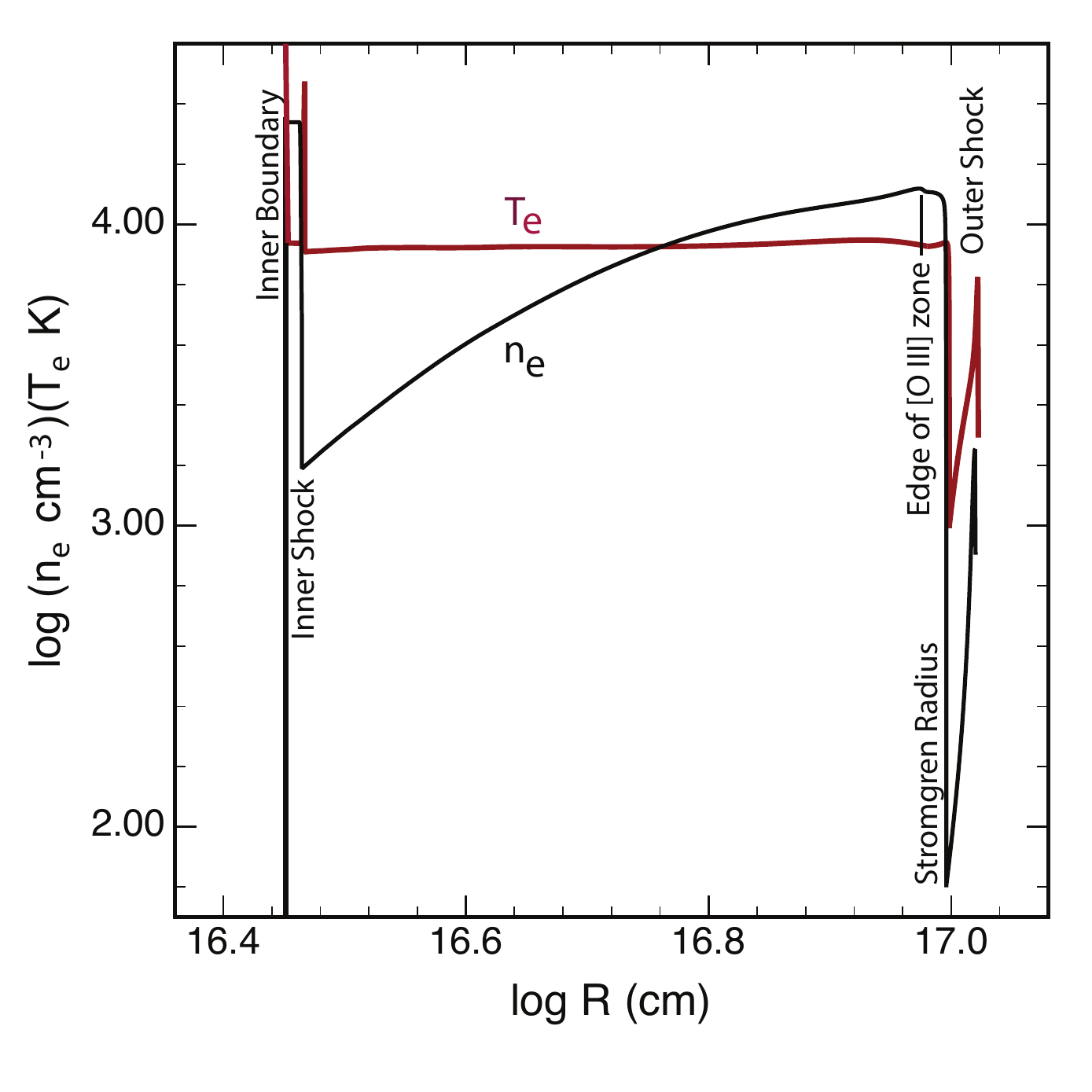}
 \caption{The run of electron temperature and density as a function of radius throughout the model. The major structural features are marked.} \label{fig:neTe}
\end{figure}

\begin{table}
 \caption{The contribution of the inner and outer shocks to some strongly affected nebular line intensities as predicted by the model, all scaled to I(H$\beta$)=100.}\label{table:shocks}
 \scalebox{0.8}{
\begin{tabular}{lccccc}
\hline
Lambda (\AA)	& Ion 	& Nebula	& Inner Shell & Outer Shock & Sum \\
\hline
3727,9	& [O II]	& 171.2 & 15.0 & 4.2 & 168.3\\
3867		& [Ne III] 	& 2.4	     & 65.0  & 0.0	 &    2.9 \\
4068	& [S II]	& 1.9     & 0.04  & 2.8 &   1.9 \\
4363	& [O III]	& 0.87   & 7.2    & 0.0 & 0.91 \\
4711	& [Ar IV] 	& 0.41  & 0.86	& 0.0 & 0.41 \\
4861	& H$\beta$ & 100   & 100  & 100 & 100 \\
5007	& [O III]	& 197.6  & 1114 & 0.0 & 202.9 \\
5200	& [N I]     	& 0.24  & 0.0  & 21.3  & 0.45 \\
6300	& [O I]	& 1.72 & 0.0  & 60.0  & 2.28 \\
Fractional & & 0.9824 & 0.0100 & 0.0076 & 1.000\\
 Flux (H$\beta$ =1): & & & & \\
& & & & & \\
\hline			
\end{tabular}} \\
\end{table}

\section{Results of the model}
\subsection{Emission Line Intensities}
Our model provides predictions for a total of 154 emission lines in the optical spectrum of IC 418. For the permitted lines of heavy elements, the model only accounts for the recombination contribution to the line. As \citet{Escalante12} demonstrated, the intensity of most of these lines are dominated by fluorescence. We have therefore used the ratio of fluorescence to recombination computed by these authors to correct our predictions for these lines. In Table \ref{table2} we present the list of modelled emission lines with their predicted fluxes relative to H$\beta$, the measured de-reddened line intensities and errors, as well as the de-reddened line intensities given by \citet{Sharpee03} from their \'echelle observations. The lines marked with an asterix (*) have been corrected for the fluorescent process.

Overall the fit of the model with the observations is very good, with most of the predicted line strengths falling within $\pm20$\% of the observed value. Strong lines which are poorly modelled are [Ne III] ($\sim 25$\% low) and [S III] $\lambda6312$ (overestimated by a factor of two -- but this line is very sensitive to the electron temperature). The [Fe II]/[Fe III] ratios are poorly reproduced -- but this ratio is very sensitive to the adopted charge-exchange reaction rates.

\begin{table*}
\centering \caption{A comparison of the observed line fluxes, the predictions of the model, and the line fluxes of \citet{Sharpee03}. }\label{table2}
 \scalebox{0.80}{
\begin{tabular}{clcccclccc}
 \hline
Wavelength  & ID      & Observed Flux            & Model  & Sharpee03 &   Wavelength & ID      & Observed  Flux         & Model  & Sharpee03 \\
(\AA)              &            & (H$\beta$=100)  &               &                       & (\AA)      &                          & (H$\beta$=100)             & \\
\hline
3613.642 & He I      & $0.5110 \pm 0.0242$ & 0.4892  & 0.5577   &   4890.856 & O II      & $0.0100 \pm 0.0004$ & 0.0030* & 0.0137   \\
3697.158 & H I       & $1.4540 \pm 0.1274$ & 1.1670  & 1.6649   &   4921.931 & He I      & $1.0721 \pm 0.0186$ & 1.2085 & 1.2186    \\
3703.859 & H I       & $1.7290 \pm 0.3146$ & 1.3710  & 1.9143   &   4931.227 &[O III],[Fe III]& $0.0321 \pm 0.0007$ & 0.0270 & 0.0294 \\
3711.977 & H I       & $2.0020 \pm 0.0860$ & 1.6340  & 2.2926   &   4958.911 & [O III]   & $64.179 \pm 1.0240$ & 71.2760 & 72.7233  \\
3721.750 & H I,[S III]& $3.0970 \pm 0.1425$ & 3.7458  & 3.3082  &   4987.210 &  [Fe III ]& $0.0154 \pm 0.0008$ & 0.0140 & 0.0165    \\
3726.032 & [O  II ]   & $126.580 \pm 8.984$ & 124.316 & 123.7908&   4906.830 & O II      & $0.0050 \pm 0.0003$ & 0.0045* &          \\
3728.815 & [O  II ]   & $52.580 \pm 10.901$ & 51.6340 & 52.3426 &   4924.529 & O II      & $0.0134 \pm 0.0081$ & 0.0070* &          \\
3734.375 & H I       & $2.7380 \pm 0.1292$ & 2.4490  & 3.0689   &   4994.360 & N II      & $0.0310 \pm 0.0008$ & 0.0313* & 0.0423   \\
3750.158 & H I       & $3.5430 \pm 0.1603$ & 3.0910  & 4.0585   &   5001.480 & N II      & $0.0327 \pm 0.0019$ & 0.0366* &          \\
3770.637 & H I       & $4.4060 \pm 0.1952$ & 3.9970  & 4.0585   &   5006.843 & [O III]   & $199.430\pm 5.017$ & 206.0137 & 214.935  \\
3777.134 & Ne II     & $0.0053 \pm 0.0010$ & 0.0062  & 0.0050   &   5015.678 & He I      & $2.2745 \pm 0.0370$ & 2.3269 & 2.3922    \\
3797.904 & H I       & $5.5240 \pm 0.2487$ & 5.3170  & 5.6643   &   5047.739 & He I      & $0.1890 \pm 0.0032$ & 0.1685 & 0.1887    \\
3835.391 & H I       & $7.3140 \pm 0.3354$ & 7.3130  & 9.4921   &   5158.792 & [Fe II]   & $0.0080 \pm 0.0005$ & 0.0041 & 0.0102    \\
3864.431 & O II      & $0.0067 \pm 0.0033$ & 0.0021* &          &   5191.822 & [Ar III]  & $0.0326 \pm 0.0007$ & 0.0400 & 0.0386    \\
3868.764 & Ne III ]   & $4.3060 \pm 0.2226$ & 3.0320  & 3.0916  &   5197.902 & [N I ]    & $0.2818 \pm 0.0036$ & 0.2568 & 0.2011    \\
3882.194 & O II      & $0.0071 \pm 0.0015$ & 0.0100*  & 0.0063   &   5200.257 & [N I]     & $0.1547 \pm 0.0022$ & 0.3560 & 0.1173    \\
3888.800 & H I, He I  & $14.400 \pm 1.071 $ & 19.6500 & 16.0294 &   5261.633 & [Fe II]   & $0.0059 \pm 0.0005$ & 0.0017 & 0.0062    \\
3933.663 & Ca II     & $0.0042 \pm 0.0005$ & 0.0064  &          &  5270.403 & [Fe III]  & $0.0231 \pm 0.0006$ & 0.0590 & 0.0151     \\
3888.800 & H I , He I & $14.400 \pm 1.0706$ & 19.6500 & 16.0294 &   5273.364 & [Fe II]   & $0.0030 \pm 0.0004$ & 0.0011 & 0.0018    \\
3964.729 & He I      & $0.8235 \pm 0.0509$ & 1.0545  & 0.9220   &   5452.080 & N II      & $0.0019 \pm 0.0005$ & 0.0115* & 0.0018   \\
3967.471 & [Ne III ]  & $1.3401 \pm 0.1093$ & 0.9133  & 0.9751  &   5462.590 & N II      & $0.0042 \pm 0.0004$ & 0.0105* & 0.0042   \\
3970.079 & H I       & $16.014 \pm 0.5726$ & 15.8730 & 16.8492  &   5478.086 & N II      & $0.0025 \pm 0.0002$ & 0.0024* & 0.003    \\
4026.209 & He I      & $1.8894 \pm 0.0511$ & 2.1130  & 2.0978   &   5480.060 & N II      & $0.0053 \pm 0.0003$ & 0.0136* & 0.0056   \\
4035.080 & N II      & $0.0023 \pm 0.0005$ & 0.0033  & 0.0071   &   5517.709 & [Cl III]  & $0.1792 \pm 0.0033$ & 0.1995 & 0.1819    \\
4041.310 & N II      & $0.0059 \pm 0.0014$ & 0.0070  & 0.0121   &   5537.873 & [Cl III ] & $0.3651 \pm 0.0070$ & 0.4104 & 0.3560    \\
4068.600 & [S II],O II& $2.3690 \pm 0.0780$ & 1.9391  & 1.8077  &   5577.339 & [O I]     & $0.0308 \pm 0.0010$ & 0.0130 & 0.0263    \\
4072.152 & O  II     & $0.0333 \pm 0.0105$ & 0.0359  & 0.0327*   &   5666.630 & N II      & $0.0407 \pm 0.0020$ & 0.0384* & 0.0414   \\
4076.349 & [S II]    & $0.8679 \pm 0.0243$ & 0.6228  & 0.7653   &   5676.020 & N II      & $0.0191 \pm 0.0094$ & 0.0147* & 0.0197   \\
4087.153 & O II      & $0.0057 \pm 0.0009$ & 0.0034  & 0.0045   &   5679.560 & N II      & $0.0602 \pm 0.0022$ & 0.0661* & 0.0674   \\
4089.288 & O II      & $0.0150 \pm 0.0012$ & 0.0150  & 0.0114   &   5686.210 & N II      & $0.0112 \pm 0.0007$ & 0.0087* & 0.0127   \\
4095.644 & O II      & $0.0203 \pm 0.0083$ & 0.0025  & 0.0042   &   5710.780 & N II      & $0.0150 \pm 0.0011$ & 0.0140* & 0.0136   \\
4097.257 & O II      & $0.0063 \pm 0.0083$ & 0.0096* & 0.0115   &   5754.595 & [N II]    & $2.7304 \pm 0.0902$ & 2.2960 & 2.7615    \\
4101.742 & H I       & $26.6525\pm 0.9271$ & 25.8480 & 24.8041  &   5875.664 & He I     & $12.4705 \pm 0.4257$ & 12.2260 & 13.6746  \\
4104.723 & O II      & $0.0243 \pm 0.0106$ & 0.0090* & 0.0160   &   5927.810 & N II      & $0.0160 \pm 0.0009$ & 0.0313* & 0.0191   \\
4110.786 & O II      & $0.0111 \pm 0.0019$ & 0.0032* & 0.0075   &   5931.790 & N II      & $0.0215 \pm 0.0015$ & 0.0112* & 0.0270   \\
4120.835 & He I      & $0.2022 \pm 0.0069$ & 0.1707  & 0.2062   &   5941.650 & N II      & $0.0273 \pm 0.0034$ & 0.0142* & 0.0315   \\
4132.800 & O II      & $0.0116 \pm 0.0016$ & 0.0099* & 0.0083   &   5952.390 & N II      & $0.0071 \pm 0.0006$ & 0.0074* & 0.0052   \\
4153.298 & O II      & $0.0171 \pm 0.0014$ & 0.0129* & 0.0184   &   6149.298 & C II      & $0.0253 \pm 0.0008$ & 0.0150 & 0.0253    \\
4267.140 & C II      & $0.4896 \pm 0.0115$ & 0.3520  & 0.5712   &   6300.304 & [O I]     & $2.8345 \pm 0.0976$ & 2.4365 & 2.1753    \\
4275.551 & O II      & $0.0052 \pm 0.0003$ & 0.0068 & 0.0065    &   6312.063 & [S III]   & $0.8527 \pm 0.0277$ & 1.6550 & 0.8566    \\
4277.894 & O II      & $0.0006 \pm 0.0002$ & 0.0012 & 0.0033    &   6363.776 & [O I]     & $0.9415 \pm 0.0331$ & 0.7793 & 0.7594    \\
4294.700 & O II      & $0.0030 \pm 0.0005$ & 0.0029 & 0.0059    &   6527.231 & [N II]    & $0.0290 \pm 0.0015$ & 0.0280 & 0.0285    \\
4303.823 & O II      & $0.0078 \pm 0.0007$ & 0.0007 & 0.0066    &   6548.052 & [N II]    & $49.2466 \pm 1.5995$& 52.3040 & 53.6007  \\
4307.232 & O II      & $0.0129 \pm 0.0010$ & 0.0013 & 0.0119    &   6562.819 & H I      & $283.3268 \pm 9.7787$& 287.0080 & 312.043 \\
4315.360 & O II      & $0.0020 \pm 0.0005$ & 0.0006 &           &   6578.053 & He I,C II & $0.5215 \pm 0.0243$ & 0.5316 & 0.5374    \\
4332.694 & O II      & $0.0018 \pm 0.0008$ & 0.0021* &          &   6583.454 & [N  II]  & $147.2072 \pm 5.0507$ &153.8852 &162.9287 \\
4340.471 & H I       & $48.0351\pm 1.0194$ & 46.7602 & 44.8053  &   6678.152 & He I      & $3.2936 \pm 0.1118$ & 3.4798 & 3.8721    \\
4363.209 & [O III]   & $0.7882 \pm 0.0194$ & 0.9280 & 0.9353    &   6716.440 & [S II]    & $2.2171 \pm 0.0692$ & 2.1834 & 2.0831    \\
4387.929 & He I      & $0.4925 \pm 0.0117$ & 0.5603 & 0.5462    &   6730.816 & [S II]    & $4.5866 \pm 0.1622$ & 4.2908 & 4.4215    \\
4437.553 & He I      & $0.0716 \pm 0.0019$ & 0.0696 & 0.0792    &   7135.792 & [Ar III]  & $7.0589 \pm 0.2750$ & 7.4260 & 8.2608    \\
4471.502 & He I      & $3.9804 \pm 0.0924$ & 4.1424 & 4.4921    &   7231.327 & C II      & $0.1939 \pm 0.0100$ & 0.4112* & 0.1692   \\
4491.222 & O II      & $0.0119 \pm 0.0008$ & 0.0021 & 0.0125    &   7236.416 & C II      & $0.5055 \pm 0.0230$ & 0.4915* & 0.4673   \\
4566.837 & Mg I]     & $0.5226 \pm 0.0124$ & 0.5220 & 0.4291    &   7281.351 & He I      & $0.7404 \pm 0.0221$ & 0.6080 & 0.7911    \\
4601.478 & N II      & $0.0232 \pm 0.0009$ & 0.0209* & 0.0263   &   7291.469 & Ca II     & $0.0149 \pm 0.0010$ & 0.0152 & 0.0123    \\
4607.100 & N II      & $0.0235 \pm 0.0008$ & 0.0206* & 0.0257   &   7318.923 & [O II]    & $3.7000 \pm 0.3500$ & 3.5080 & 3.6886    \\
4613.868 & N II      & $0.0159 \pm 0.0008$ & 0.0131* & 0.0182   &   7319.989 & [O II]    & $10.7196 \pm 0.8900$ & 10.4340 & 10.0586 \\
4621.390 & N II      & $0.0211 \pm 0.0008$ & 0.0262* & 0.0264   &   7329.665 & [O II]    & $6.5190 \pm 0.3530$ & 5.8160 & 5.8617    \\
4630.540 & N II      & $0.0683 \pm 0.0019$ & 0.0737* & 0.0805   &   7330.735 & [O II]    & $5.7000 \pm 0.3300$ & 5.5320 & 5.6377    \\
4643.086 & N II      & $0.0298 \pm 0.0012$ & 0.0281* & 0.0344   &   7377.829 & [Ni II]   & $0.0042 \pm 0.0023$ & 0.0042 & 0.0049    \\
4658.051 & [Fe III]  & $0.0035 \pm 0.0020$ & 0.0689 & 0.0274    &   7751.109 & [Ar III]  & $1.8159 \pm 0.0788$ & 1.7879 & 2.1967    \\
4667.010 & [Fe III]  & $0.0035 \pm 0.0005$ & 0.0038 & 0.0028    &   7771.944 & O  I      & $0.0282 \pm 0.0012$ & 0.0162 & 0.0352    \\
4699.218 & O II      & $0.0097 \pm 0.0097$ & 0.0183* & 0.0133   &   7774.166 & O  I      & $0.0195 \pm 0.0009$ & 0.0162 & 0.0215    \\
4701.535 & [Fe III]  & $0.0162 \pm 0.0006$ & 0.0320 &           &   7775.388 & O  I      & $0.0115 \pm 0.0005$ & 0.0162 & 0.0130    \\
4705.346 & O II      & $0.0147 \pm 0.0005$ & 0.0214 & 0.0183    &   8433.661 & [Cl III]  & $0.0095 \pm 0.0014$ & 0.0110 & 0.0062    \\
4713.171 &  He I     & $0.5881 \pm 0.0144$ & 0.4445 & 0.6098    &   8437.955 & H I       & $0.4057 \pm 0.0153$ & 0.3370 & 0.4345    \\
4733.906 & [Fe III]  & $0.0067 \pm 0.0005$ & 0.0140 & 0.0048    &   8467.254 & H I       & $0.4645 \pm 0.0210$ & 0.3940 & 0.5050    \\
4740.123 & [Ar IV]   & $0.0040 \pm 0.0005$ & 0.0044 & 0.0036    &   8480.859 & [Cl III]  & $0.0113 \pm 0.0007$ & 0.0110 & 0.0108    \\
4754.687 & [Fe III]  & $0.0080 \pm 0.0005$ & 0.0130 & 0.0047    &   8502.483 & H I       & $0.5459 \pm 0.0202$ & 0.4660 & 0.5983    \\
4769.431 & [Fe III]  & $0.0039 \pm 0.0004$ & 0.0110 & 0.0062    &   8545.383 & H I       & $0.6407 \pm 0.0241$ & 0.5600 & 0.6807    \\
4779.720 & N II      & $0.0151 \pm 0.0006$ & 0.0305* & 0.0179   &   8616.950 & [Fe II]   & $0.0096 \pm 0.0007$ & 0.0105 & 0.0094    \\
4788.130 & N II      & $0.0166 \pm 0.0007$ & 0.0185* & 0.0195   &   8578.697 & [Cl II]   & $0.3165 \pm 0.0119$ & 0.1927 & 0.2844    \\
4803.290 & N II      & $0.0176 \pm 0.0009$ & 0.0102* & 0.0261   &   8598.392 & H I       & $0.7708 \pm 0.0284$ & 0.6840 & 0.8362    \\
4810.299 & N II      & $0.0026 \pm 0.0006$ & 0.0035* & 0.004    &   8665.018 & H I       & $0.9290 \pm 0.0349$ & 0.8500 & 0.9501    \\
4814.544 & [Fe II]   & $0.0040 \pm 0.0005$ & 0.0013 & 0.008     &   8727.126 & [C I]     & $0.0373 \pm 0.0016$ & 0.0393 & 0.0334    \\
4861.333 & H I       & $100.000 \pm 0.480$ & 100.000 & 100.000  &   8750.472 & H  I      & $1.1829 \pm 0.0437$ & 1.0780 & 1.3112    \\
4880.996 & [Fe III]  & $0.0231 \pm 0.0012$ & 0.0760 & 0.0154    &   8829.391 & [S III]   & $0.0059 \pm 0.0004$ & 0.0110 & 0.0048    \\
4889.623 & [Fe II]   & $0.0046 \pm 0.0005$ & 0.0011 &           &   8862.783 & H I       & $1.5432 \pm 0.0588$ & 1.3990 & 1.6217    \\
 \hline

\end{tabular}}
\end{table*}

In Figure \ref{fig7} we show the fit of the model versus the observations. Perfect agreement is represented by the central solid line, while the lines on each size represent a difference of $\pm 20$\% in the relative flux.
\begin{figure}
 \includegraphics[width=\columnwidth]{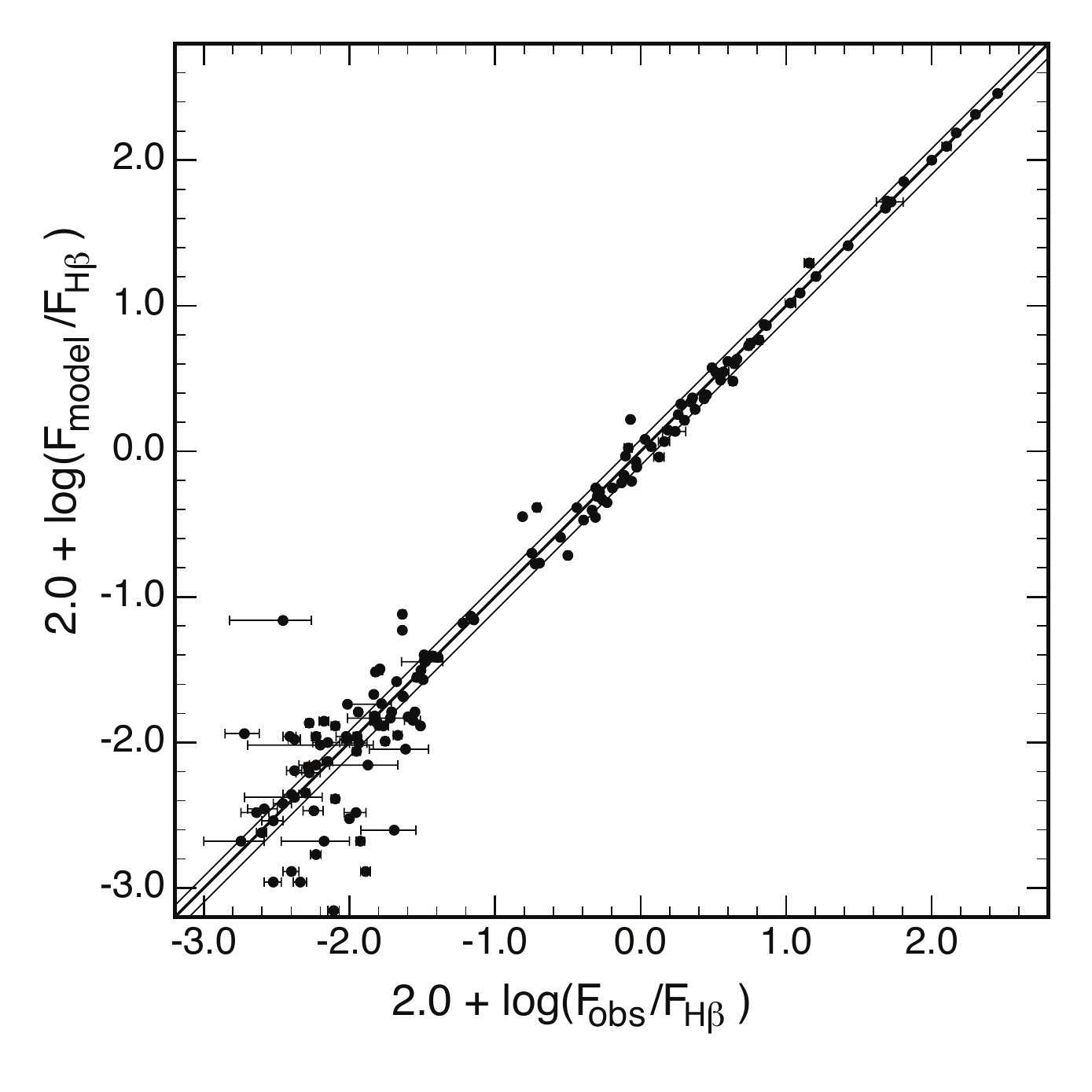}
 \caption{The fit of the model versus the (de-reddened) observations of IC 418. The dynamic range of the fitted intensities is $\sim 10^5$. Apart from some faint recombination lines, the quality of the fit for individual lines is generally better than $\pm 20$\%.} \label{fig7}
\end{figure}

It is of interest to compare our measured line intensities with those given by \citet{Sharpee03}. These observations were made at much higher resolution than ours, and the integration times were much longer, which facilitates the detection of fainter lines. Against this, however, is the notorious difficulty of flux calibrating \'echelle data due to the strong variation in grating efficiency as one goes off-blaze in each order. Furthermore, the \citet{Sharpee03} data covers only a portion of the nebula, which could cause problems associated with sampling the excitation structure of the nebula. 

In Figure \ref{fig8} we compare our own de-reddened data with the de-reddened fluxes given by \citet{Sharpee03} for the same lines we used in the model fit. Perfect agreement is represented by the thick solid line, while the thinner lines on each size represent a difference of $\pm 20$\%. The agreement between these two independent data sets is  remarkable, and bears testament  to the both the quality of the observations and data reduction procedures. Furthermore, it is clear that the observing strategy of \citet{Sharpee03} enabled them to obtain a good approximation to the integral spectrum of the nebula.
\begin{figure}
 \includegraphics[width=\columnwidth]{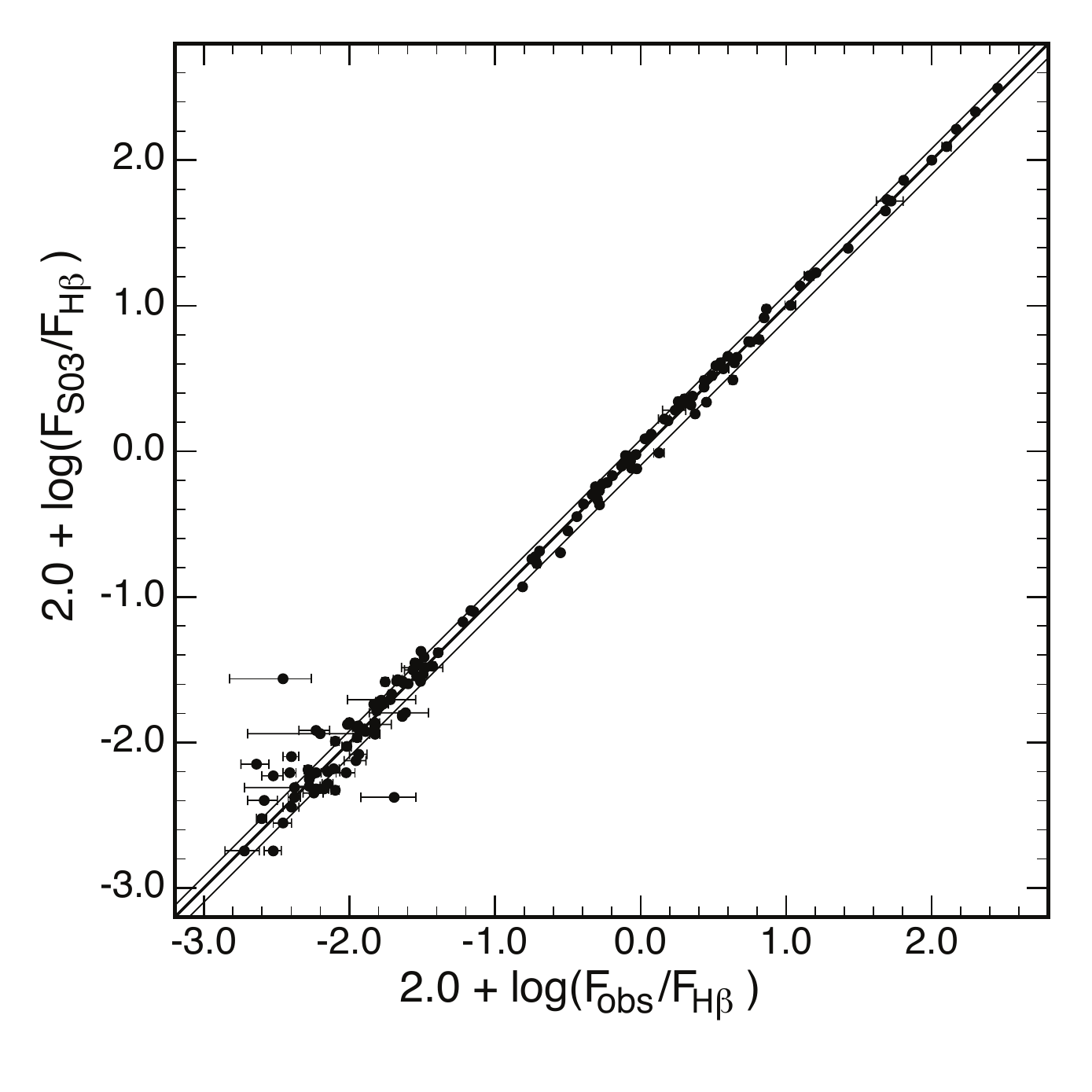}
 \caption{The fit of the observed (de-reddened) line fluxes versus the de-reddened observations by \citet{Sharpee03} (S03) of these same lines in IC 418.  The scatter between these two data sets is generally less than $\pm 20$\% over the full dynamic range.} \label{fig8}
\end{figure}

\subsection{The Nebular Continuum}
The fit of the measured (de-reddened) nebular continuum (black points) with the theoretical (continuum + emission line) spectrum of the PNe (blue line) is shown in Figure \ref{fig9}. It is clear that the theoretical continuum provides a good description of the observations except in the region $\lambda\lambda 3650-4050$\AA. This difference is due to either mis-measurement of the underlying continuum in the presence of many faint overlapping lines -- many lines are fit with a fixed width in this wavelength region -- or else scattering of the stellar UV by dust in and around the nebula. The size of both the Paschen and Balmer jumps are well simulated, as is the overall slope of the continuum, indicating that the mean nebular temperature characterising the model is close to that  prevailing in the nebula itself.
\begin{figure*}
 \includegraphics[scale=0.6]{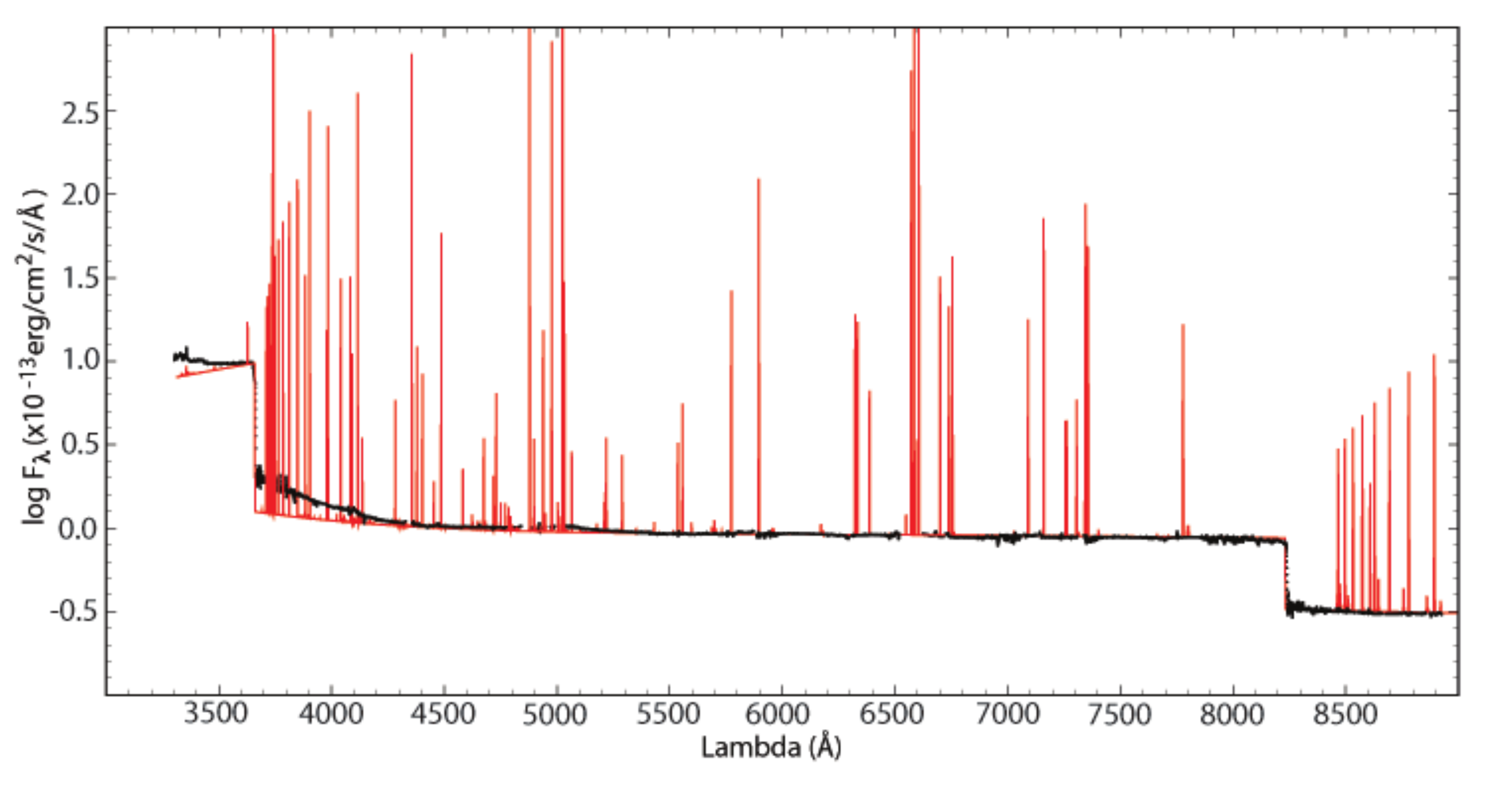}
 \caption{The fit of the model spectrum (in red) with the (de-reddened) observations of the continuum in IC 418 (in black). The fitted emission lines have been removed from this spectrum. The rise in the apparent continuum below 4000\AA\ is largely due to the mis-measurement of the underlying continuum in the presence of many faint overlapping lines. There may also be scattered light resulting from dust scattering of the stellar UV continuum.} \label{fig9}
\end{figure*}

\subsection{Chemical Abundances} \label{Sec6.3}
The abundances derived in the model are listed in Table \ref{table6}. Here we give separately the messured gas phase abundances, and the implied total abundances using the dust depletion pattern for the local interstellar medium from \citet{Jenkins09} for a base $\log{\mathrm Fe / H}$ depletion of -2.25. Clearly, we do not expect the dust in IC 418 to share this depletion pattern, so the total (Gas + Dust) abundances given in the table should be treated as indicative only. Table \ref{table6} also compares our results with the gas phase abundances derived by \citet{Morisset09}. This paper also presents the comparison with earlier work. Considering that we are using an independent data set, different analysis techniques, and different modelling codes, we find that the mutual agreement for the important coolants is very good, generally better than 0.1\,dex.

As described above, we have used the Nebular Empirical Abundance Tool (NEAT; \citet{Wesson12} to derive the electron temperature and density from the low and medium-ionization zones. In Table \ref{table6}, we also present the elemental abundances of  nitrogen, oxygen, neon, argon, sulfur, and chlorine that determined from collisionally excited lines (CELs) and those of hydrogen, helium, carbon derived from optical recombination lines (ORLs). The ionization correction factors of \citet{Delgado-Inglada14} were used to correct for unseen ions. The chemical abundances are found to be in agreement with that of \citet{Delgado-Inglada15} and \citet{Pottasch04}. However, in general, the chemical abundances are generally systematically lower than those derived by the detailed modelling presented here, and those derived in a similar manner by \citet{Morisset09}. This illustrates once again the perennial problem of differences between $T_e$--derived abundances, and those delivered by detailed photoionisation modelling. Abundances delivered by $T_e+$ICF calibration are systematically lower than those used in the photoionisation models. This is mainly caused by regions of higher than average temperature over-weighting temperature-sensitive lines such as [O III] $\lambda4363$. The ICFs do not seem to be playing a major role, since the computed ICFs agree are in broad agreement with the photoionisation model (for OI:OII:OII we have the ratios 0.024\,:\,0.490\,:\,0.486 from the ICFs given by NEAT, while the ionic column densities  from the model are in the ratio 0.018\,:\,0.496\,:\,0.486.  At least part of the disagreement between the two techniques is due to real temperature gradients we have demonstrated to exist in IC 418 (c.f. Figure \ref{fig5}). These large-scale temperature gradients play a role analogous to the smaller-scale temperature fluctuations first introduced by \citet{Peimbert67} and used by very many others since, \emph{e.g.} \citet{Esteban02,Garcia07,Pena12}. In particular, \citet{Kingdon95} has attempted to reproduce temperature fluctuations in the context of temperature variations produced in detailed phototionisation models.

\begin{table}
 \caption{The measured gas phase and implied total (Gas+ Dust) abundances. The gas phase abundances may be compared with the \citet{Morisset09} abundances (MG09), and with the abundances derived from the NEAT code.}\label{table6}
 \scalebox{1.0}{
\begin{tabular}{lccccc}
 \hline
        & \multicolumn{4}{|c|}{$12 + \rm {\log(X/H)}$}\\\cline{2-4}
        &                  &     \\
Element &   Gas Phase   &   Total   & MG09  & NEAT \\
\hline
  H     &   12.00  &   12.00  &  12.00 & 12.00\\               
  He    &   11.04      &   11.04    & 11.08 & 10.90 \\       
  C     &   8.71   &   8.92  & 8.90 & 8.74 \\              
  N     &   7.98   &   8.08   & 8.00 & 7.65 \\             
  O     &   8.62   &   8.86   & 8.60 & 8.34\\             
  Ne    &   8.14   &   8.14  & 8.00 & 7.50 \\              
  Mg    &   7.04    &   8.32  &  7.05 &  --- \\                
  Si    &   6.84   &   7.50   & 7.10 &  --- \\             
  S     &   6.89   &   6.89   & 6.65 & 6.28\\             
  Cl    &   4.89    &   5.70 & 5.00 & 4.95 \\              
  Ar    &   6.08   &   6.08   & 6.20 & 6.23 \\             
  Ca    &   3.84   &   6.99   & --- &  ---\\
  Fe    &   4.97   &   7.22   & 4.60 &  --- \\             
  Ni    &   4.02    &   6.46   & --- &  --- \\
\hline
\end{tabular}} \\
\end{table}

The fluorescent corrections given by \citet{Escalante12} work rather well to remove the abundance discrepancy, which is often inferred to exist in other PNe, between the permitted and the forbidden lines. In Figure \ref{fig10} we show the comparison of the model predictions for the forbidden lines (filled circles) and the fluorescent - corrected ``recombination" lines of N II and O II (open circles). The error bars refer to the line measurement errors only.

\begin{figure*}
 \includegraphics[scale=0.55]{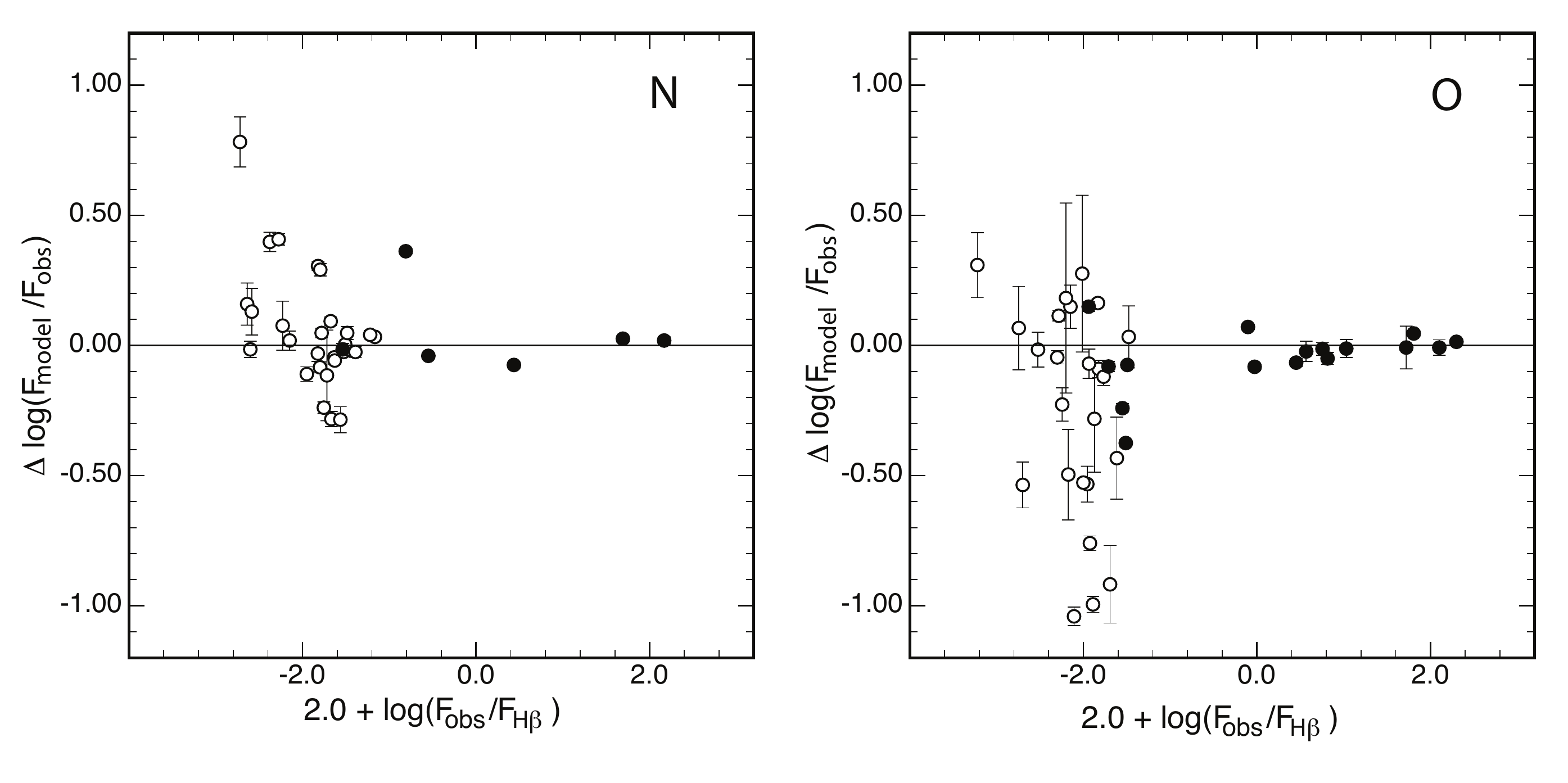}
 \caption{A comparison of the IC 418 model fit for the fluorescence - corrected permitted lines of N II and O II (open circles) and the corresponding fit for the forbidden lines of any ionisation stage (filled circles). The effect of the fluorescence corrections given by \citet{Escalante12} are to remove the abundance discrepancy between permitted and forbidden lines that would otherwise have been present.} \label{fig10}
\end{figure*}

\subsection{Comparison with the UV and IR spectrum}
Although we have not specifically modelled the UV and IR spectrum of IC 418, nonetheless it is of interest to see how well the model performs in reproducing the de-reddened UV and IR line intensities given by \citet{Pottasch04} and by \citet{Liu01}, who used the LWS on the ISO observatory to measure the longer wavelength (43-198$\mu$m) lines. The comparison between the model and these observations is given in Table \ref{table7}. Overall, for the IR lines,  the fit is satisfactory -- with the exception of [O I] and [C II] lines the L1-norm is 0.12. For the UV lines (with their larger and more uncertain reddening corrections) the L1-norm is 0.17. All the oxygen lines fit the observations very closely.

As \citet{Liu01} point out, IC 418 is blessed with an exceptionally strong photo-dissociation region (PDR) which provides a large contribution to the observed intensity of the [O I]) and [C II] lines. However, since the MAPPINGS code does no include a molecular formation/destruction chemical matrix, but only deals with atomic phases, we cannot model the temperature, density and molecular / ion balance in PDRs with any accuracy. Any prediction of the [O\,I] and [C\,II] lines would therefore be completely unreliable. This is clear from Table \ref{table7}. All we can do with reasonable accuracy is to compute the dust temperature distribution in these regions.

The 35.81$\mu$m [Si II] line is the only line of this element observed, and this has been used to determine the Si abundance shown in Table \ref{table7}, although this line might be dominated by the contribution of emission from the PDR. The Mg abundance was obtained from only one line in the optical spectrum, so it is pleasing to see how well the modelled Mg II 2798\AA\ line matches the observations. The carbon lines provide a useful further constraint on the abundance estimated from the [C I] $\lambda 8727$ line, and from the four CII permitted lines which are observed in the optical.

\begin{table}
 \caption{The comparison of the model with UV and IR lines observed by \citet{Pottasch04} and \citet{Liu01} in IC 418. Line fluxes are relative to H$\beta = 100$. }\label{table7}
 \scalebox{1.0}{
\begin{tabular}{lccc}
\hline
Lambda (\AA)	& Ion 	&  Flux	& Model \\
\hline
1663	& O III]	& $<0.7$	& 0.71 \\
1750	& N III]	& 0.58	& 0.87 \\
1909	& C III]	& 27.59	& 49.3 \\
2325	& C II]	& 81.2	& 44.5 \\
2471	& [O II]	& 19.2	& 19.8 \\
2798	& Mg II	& 16.3	& 21.57 \\
& & & \\
\hline			
Lambda($\mu$m)	& Ion 	& Flux	& Model \\
\hline			
2.625	& H I	         & 4.7	   & 4.68  \\
4.052	& H I	         & 8.6	   & 8.21  \\
7.46	          & H I	         & 2.7	   & 2.61  \\
8.99	          & [Ar III]	& 7.3	   & 7.39  \\
10.51	& [S IV]	& 1.3	   & 2.05  \\
12.81	& [Ne II]	& 53.1 & 82.7  \\
15.55	& [Ne III]	& 9.5	   & 4.23  \\
18.71	& [S III]	& 15.2 & 24.4  \\
33.47	& [S III]	& 2.6	   & 4.31  \\
34.81	& [Si II]	& 0.9	   & 0.90  \\
51.81	& [O III]	& 14.7 & 14.6  \\
57.34	& [N III]	& 3.2   & 3.2 \\
63.10	& [O I]	& 10.1 & 0.34 \\
88.36	& [O III]	& 2.4	   & 1.95  \\
121.8	& [N II]	& 0.16 & 0.21 \\
145.6	& [O I] 	& 0.10 & 0.24 \\
157.7	& [C II]	& 0.99 & 0.37 \\
\hline
\end{tabular}} \\
\end{table}

\subsection{Surface Brightness Distribution}
As described above, the radiation pressure induces a steep density gradient in the ionised plasma, which leads to an increase in the projected surface brightness with radius. In addition, the inner shock contributes appreciably to the H$\alpha$ surface brightness. On the basis of our idealised spherical model, we have computed the surface brightness as a function of radius. This is shown in Figure \ref{fig11}. 
\begin{figure}
 \includegraphics[width=\columnwidth]{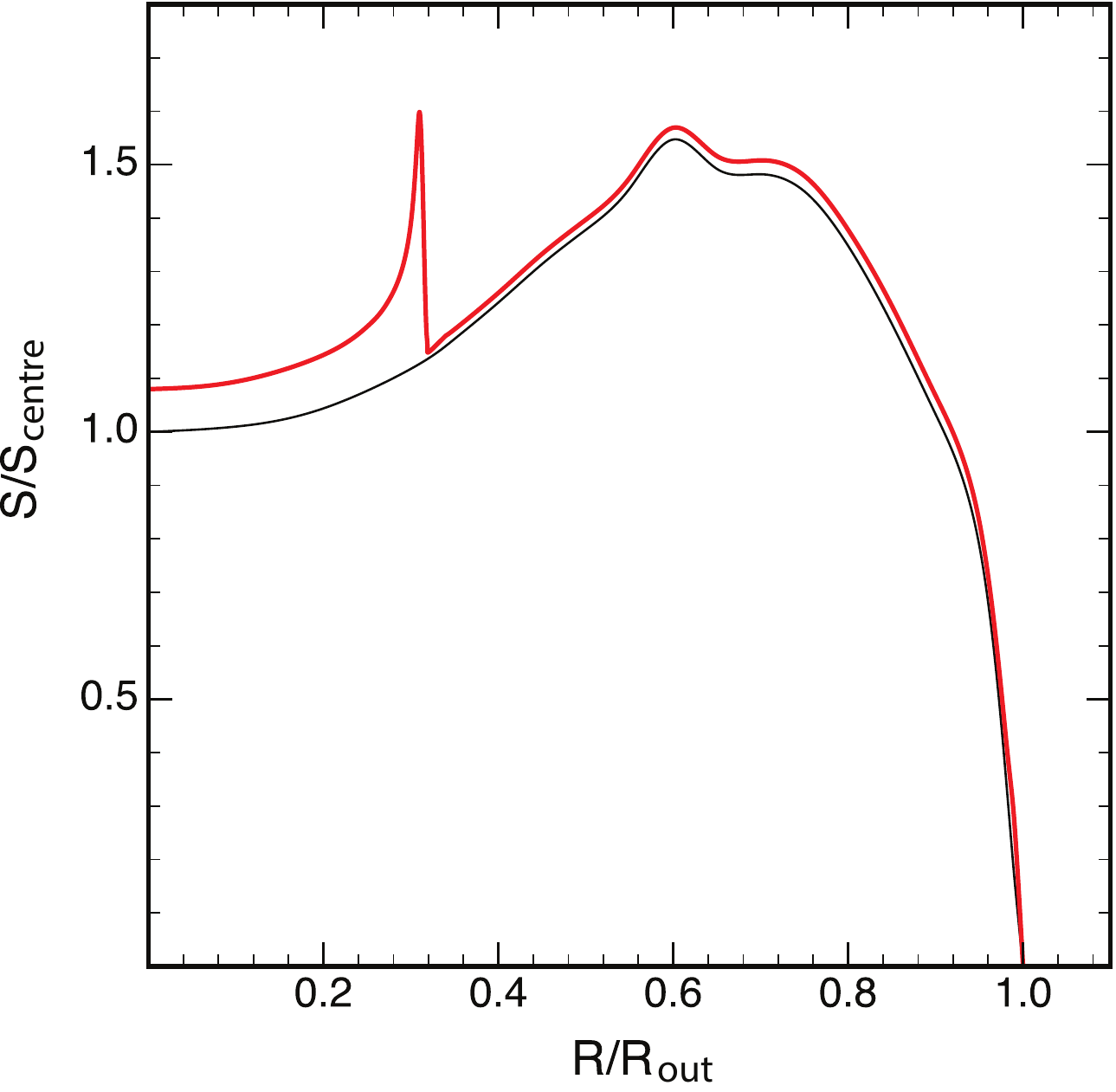}
 \caption{The computed brightness distribution for IC 418. The black curve is for the photoionised nebula only, while the red curve shows how this is modified by the presence of the two shocks.} \label{fig11}
\end{figure}

This surface brightness distribution can be compared with that shown by \citet{Morisset09}. The inner shock appears sharper here because it is represented by a smooth spherical shock, rather than the more filamentary structure seen in the HST data.

\section{Discussion \& Conclusions}
We have presented a physical model for IC 418 which provides a good description of the emission line spectrum, while at the same time matching the density sensitive line ratios, the inferred absolute H$\beta$ flux and the observed angular diameter of the nebula. From the photoionisation model we estimate a distance of $1.0\pm0.1$\,kpc, slightly less than the value usually adopted. This result is due to the strong competition of dust for the ionising photons in a nebula in which radiation pressure exceeds the static gas pressure. The absorption of the UV radiation field leads to steeply increasing pressure and density with nebular radius.

We estimate the stellar parameters to be $\log T_{\mathrm eff} = 4.525$K,  $\log L_*/L_{\odot} = 4.029$ and $\log g = 3.5$. From the hydrogen-burning \citet{VW94} evolutionary tracks, we find that this corresponds to an initial stellar mass of $\sim 2.8M_{\odot}$. This derived mass of the parent star is consistent with the mass range ($1.5-3.0M_{\odot}$) reported by \citet{Delgado-Inglada15} for the PNe which give rise to carbon-rich dust (see the introduction). Interpolating the hydrogen-burning $Z=0.016$ \citet{VW94} evolutionary tracks, we also find that the star has had only $\sim 180$\,yr since it passed through an effective temperature of  $\log T_{\mathrm eff} = 4.0$, which can be taken effectively as the time since the nebula first became ionised. Very similar results are obtained if we use the more recent models of \citet{MillerBertolami16}. For $Z=0.01$, the precursor would have had an initial mass of $2.5M_{\odot}$, while for $Z=0.02$, the estimated initial mass rises to $3.0M_{\odot}$. In both cases the transition time from an effective temperature $\log T_{\mathrm eff} = 3.8$K is $\sim100$\,yr.

Integrating the density throughout the nebula in the model, we find that the total mass inside the ionised nebula out to where hydrogen is less than 1\% ionised is $4.8\times10^{-2}M_{\odot}$. Since the nebula is composed of the gas ejected during the last gasp of the AGB stellar evolution, this mass represents the integrated mass loss over the last 2350\,yr, using the outer radius of $1.0\times10^{17}$\,cm, and taking the velocity of the AGB wind to be 13.2\,km/s from \citet{Taylor87}. From the previous paragraph, we can infer that the time that the central star actually spent on the AGB during this period was only  2270\,yr, which implies a mean AGB mass-loss rate of $2.1\times 10^{-5}M_{\odot}$/yr. This is entirely consistent with recent mass-loss rates inferred directly from the 1612 MHz circum-stellar OH maser emission for a large sample of AGB stars by \citet{Goldman17}.

The age of the nebula can also be inferred from the properties of the central mass-loss bubble. Currently, according to the model, the shock in the nebular gas is propagating at $\sim 40$\,km/s. The radius of the shocked [O III] - bright bubble is $\sim 0.01$\,pc. Using the theory of such bubbles \citep{Dyson72, Weaver77}, we derive an age of 150\,yr for the mass-loss bubble, which agrees well with the age inferred above from stellar evolutionary considerations (100--180\,yr).

Based upon the stellar wind velocity derived from the X-ray temperature of the hot inner bubble ($v_w \sim 500$\,km/s), and the pressure in the hot bubble equated to the ram pressure driving the mass-loss bubble shock in the nebular gas ($\log P_s/k = 8.4$\,cm$^{-3}$K), and assuming that the free wind region from the central star is terminated at 60\% of the radius of the inner mass-loss bubble, we infer that the current mass loss rate from the central star is $\dot M \sim 5\times 10^{-8} M_{\odot}$yr$^{-1}$, which agrees well with that obtained by \citet{Morisset09} on the basis of the absorption line profiles of the central star ($\dot M = 3.8\times 10^{-8} M_{\odot}$yr$^{-1}$).

The ionised structure appears to be very young; 150--200\,yr. This presents something of a problem, since the sound-crossing timescale of the current ionised structure is $\sim 2000$\,yr, yet the internal density distribution has already been set up as a radiation-dominated nebula. This would seem to indicate that the radiation-pressure dominated profile is actually set up during the transition from the tip of the AGB to the point where the central star started to produce ionising photons. According to the \citet{MillerBertolami16} models cited above, for $Z=0.01$, this transition time is $\sim1600$\,yr, while for $Z=0.02$, it is $\sim1200$\,yr. Both of these are comparable with the inferred sound crossing timescale.

In conclusion, we have obtained a high-quality integral field spectrum of IC 418, and have built a self-consistent model for the nebula which includes both the effects of the inner mass-loss bubble, and the outer shock in the AGB wind. Consequently we have been able to derive reliable abundances for 14 elements, which agree well with the earlier careful modelling by \citet{Morisset09}. In addition, our measured line fluxes relative to H$\beta$ agree very closely, typically within 10\%,  with those of \citet{Sharpee03}, despite the differences in the instruments used, the area over which the spectrum is integrated and the differences in both the reduction and calibration procedures. IC 418 is both dusty and carbon-rich, and its central star appears to have been a fairly massive carbon star with an initial mass in the range $2.5-3.0M_{\odot}$ and an AGB mass-loss rate of $2.1\times 10^{-5}M_{\odot}$/yr before it made its excursion across the HR Diagram to become a PNe central star with mass-loss rate of $\dot M \sim 4\times 10^{-8} M_{\odot}$/yr. 

\section*{Acknowledgements}
The authors wish to thank the referee, Christophe Morisset for his physical insights, and his careful and constructive critique, which has appreciable improved the paper. M.D. and R.S. acknowledge the support of the Australian Research Council (ARC) through Discovery project DP130103925. 

\bibliographystyle{mn2e_new}

\end{document}